\documentclass[aps,pra,showpacs,twoside,twocolumn,10pt]{revtex4-1}
\usepackage[colorlinks=true, citecolor=red, urlcolor=blue ]{hyperref}
\usepackage{times,epsfig,amssymb,amsfonts,amsmath,bm,subfigure,mathtools,amsthm,braket,soul,enumitem,color}
\usepackage[normalem]{ulem}

\newcommand{\stkout}[1]{\ifmmode\text{\sout{\ensuremath{#1}}}\else\sout{#1}\fi}
\definecolor{magenta}{rgb}{1.0, 0.0, 0.56}

\begin{document}

\title{Three qubits in less than three baths: Beyond two-body system-bath interactions\\in quantum refrigerators}
\author{Ahana Ghoshal$^1$, Sreetama Das$^{1,2,3}$, Amit Kumar Pal$^4$, Aditi Sen(De)$^1$, Ujjwal Sen$^1$}
\affiliation{$^1$Harish-Chandra Research Institute, HBNI, Chhatnag Road, Jhunsi, Allahabad 211 019, India\\
$^2$Faculty of Physics, Arnold Sommerfeld Center for Theoretical Physics, Ludwig Maximilian University Munich, Theresienstr. 37, 80333 Munich, Germany\\
$^3$Munich Center for Quantum Science and Technology (MCQST), Schellingstrasse 4, 80799 M{\"u}nchen, Germany\\
$^4$Department of Physics, Indian Institute of Technology Palakkad, Palakkad 678 557, India}

\begin{abstract}
We show that quantum absorption refrigerators, which have traditionally been studied as of three qubits, each of which is connected to a thermal reservoir, can also be constructed by using three qubits and two thermal baths, where two of the qubits, including the qubit to be locally cooled, are connected to a common bath. With a careful choice of the system, bath, and qubit-bath interaction parameters within the Born-Markov and rotating-wave approximations, one of the qubits attached to the common bath achieves a cooling in the steady state. We observe that the proposed refrigerator may also operate in a parameter regime where no or negligible steady-state cooling is achieved, but there is  considerable transient cooling. The steady-state temperature can be lowered significantly by an increase in the strength of the few-body interaction terms existing due to the use of the common bath in the refrigerator setup.
The proposed refrigerator built with three qubits and two baths is shown to provide steady-state cooling for both Markovian qubit-bath interactions between the qubits and canonical bosonic thermal reservoirs, and a simpler reset model for the qubit-bath interactions.
\end{abstract}

\maketitle

\section{Introduction}
\label{sec:intro}

The field of quantum thermodynamics~\cite{gemmer2004,*kosloff2013,*kilmovski2015,*Misra2015,*millen2016,*benenti2017,*deffner2019,vinjanampathy2016,*goold2016} has gained considerable momentum in the last two decades. It aims to understand thermodynamic principles at the quantum mechanical level~\cite{allahverdyan2000,*brandao2015,*gardas2015}, design effective quantum thermal machines~\cite{palao2001,*feldmann2003,*nimmrichter2018,*kosloff2014,*uzdin2015,*levy2012,*clivaz2019,*mitchison2019,*Bhattacharjee2020}, and explore whether these quantum machines can provide advantages over their classical counterparts~\cite{geva1992,*feldmann2000,*ying2017,*niedenzu2018,*xu2018}. These quantum machines are envisioned to aid the emergent quantum technologies via, for example, providing a better understanding of the interplay between quantum correlations and work~\cite{huber2015,*lostaglio2015,vinjanampathy2016,*goold2016}, and controlling power consumption in quantum computation (cf.~\cite{ikonen2017}). The interconnection of the subject with different fields in science, such as statistical and solid-state physics~\cite{campisi2015,*dalessio2016}, quantum information theory~\cite{gour2015,vinjanampathy2016,*goold2016}, and quantum many-body physics~\cite{dorner2012,*mehboudi2015,*reimann2015,*eisert2015,*gogolin2016,*skelt2019} has motivated researchers from different fields to explore the possibility of setting up experiments using mesoscopic systems~\cite{giazotto2006}, trapped ions~\cite{abah2012,*rossnage2016}, nuclear magnetic resonance~\cite{peterson2019}, and superconducting materials~\cite{karimi2016,*hardal2017,*manikandan2019}, where the theoretical results can be tested. 

Among the quantum thermal machines, \emph{quantum absorption refrigerators}, constituted of  quantum few-level systems, often comprising a small number of qubits~\cite{linden2010,skrzypczyk2011,*brunner2012,*brunner2014,*brask2015,correa2013,*correa2014,*silva2015,*Erdman2018,*naseem2020,mitchison2015,das2019} and/or qudits~\cite{linden2010,man2017,*friedman2019,*wang2015},  have been in focus. Among these, refrigerators made of three qubits in contact with three local thermal baths via Markovian qubit-bath interaction have attracted special interest, since no external energy is required to attain the refrigeration ~\cite{linden2010,skrzypczyk2011,*brunner2012,*brunner2014,*brask2015,correa2013,*correa2014,*silva2015,*naseem2020,mitchison2015,das2019,he2017,*du2018,*chiru2018,*seah2018,*barra2018,*hewgill2020}. In particular, these machines gather their driving energy from the heat baths, and are designed to locally cool a chosen qubit by increasing its ground-state population. A successful design of such a three-qubit quantum refrigerator would lower the temperature of the chosen qubit, often termed the \emph{cold qubit}, below its initial temperature when the system achieves a steady state during its evolution under the influences of the heat baths. It has also been shown that lower than steady-state temperature can be achieved in the transient regime by suitably tuning the system parameters~\cite{mitchison2015,das2019}. 

Besides theoretical advancements, schemes for realizing these small quantum absorption refrigerators in quantum few-level systems using quantum dots~\cite{venturelli2013}, circuit QED architectures~\cite{hofer2016}, and atom-cavity systems~\cite{mitchison2016,*mitchison2018} have also been proposed. Recently, a quantum absorption refrigerator has been implemented using three trapped ions~\cite{maslennikov2019}. These refrigerators are expected to be useful in situations where cooling of systems as small as a qubit faster than 
the equilibration time of the qubit with a heat-bath may be required \textit{in situ} and on demand, without any external energy transfer. 

The enterprise of successfully implementing these three-qubit quantum absorption refrigerators faces two major hurdles. On one hand, achieving optimal control on the performance of the refrigerator requires careful engineering of the reservoirs and their interactions with the qubits, so that the desired steady-state cooling of the chosen qubit is achieved. On the other hand, controlled attachment of a specific heat bath to an individual qubit in the three-qubit working substance is essential for the setup, such that the rest of the three-qubit system as well as the two remaining baths remain perfectly insulated from it. There have been recent efforts in developing optimal control over the reset operation of a single superconducting qubit via successful reservoir engineering~\cite{basilewitsch2019} (cf.~\cite{poyatos1996,myatt2000} and the references thereto). Also, with respect to the latter problem, the performance of a three-qubit three-bath quantum absorption refrigerator has been investigated where all the baths interact with all the qubits at different instances~\cite{Manzano2019}. However, a complete understanding of these problems with respect to small quantum absorption refrigerators is yet to be achieved. 

In this vein, we ask the following questions.
\begin{enumerate}
\item[(1)]\emph{Is it possible to set up a three-qubit two-bath quantum absorption refrigerator by connecting 
two of the three qubits
to a common thermal reservoir}?
\item[(2)]\emph{If such a refrigerator exists, what is the effect of the few-body (involving more than two bodies) interactions, resulting due to the use of the common bath, on the performance of the refrigerator}? 
\end{enumerate}
The former question is relevant from the perspective of obtaining a ``local cooling" of one of the qubits 
in situations where
attachment to a single-qubit  may not be possible for all of the available baths. Since we are interested in showing a reduction of temperature in a single qubit of a three-qubit two-bath refrigerator, we refer it as local cooling.
The precise definition of local cooling requires some notations, and is given at the end of Sec. \ref{subsubsec:two-body}.
On the other hand, the latter question arises because 
such a set-up 
potentially
gives rise to few-body interactions between the modes of the common bath and the two qubits attached to it. Moreover, answering this question helps in gaining insight on how the microscopic details of the common bath affects the performance of the refrigerator, which may lead to an efficient engineering of the bath. In this paper, we answer the first question affirmatively, and with respect to the second question, we show that the few-body interaction terms in the system-bath coupling corresponding to the common bath, aids the three-qubit two-bath refrigerator by introducing a steady-state cooling in situations where no steady-state cooling exists in the absence of the few-body interaction terms.

Specifically, in this paper, we propose a setup for a three-qubit quantum refrigerator with two thermal baths. One of the baths is attached to two of the qubits, among which one of the qubits can be cooled, thereby showing the effects of a refrigerator, while the second bath is attached to the third qubit. We consider the thermal baths to be bosonic in nature, interacting with the respective qubits via Markovian qubit-bath interactions. The interaction of the bath attached to the single qubit is considered to be governed by two-body interactions between the spin degree of freedom of the qubit and the bosonic bath modes of the thermal reservoir. On the other hand, the bath common to the  rest of the qubits results in additional contributions from a three-body and a four-body interaction term to the system-bath interaction along with the two-body interactions between the individual spin variables and the bath modes. We analytically determine the Lindblad operators corresponding to the quantum master equation governing the evolution of the system, and solve the equation numerically to determine the time dynamics of the local temperature of the individual qubits. We demonstrate that with a careful choice of the system, the bath, and the system-bath interaction parameters within the Born-Markov and the rotating wave approximations, one of the qubits attached to the common bath undergoes a ``local" steady-state cooling.  

Our investigation reveals that the three-qubit two-bath refrigerator can also operate in a region of the parameter space where little or no steady-state cooling is obtained.  In these situations, the steady-state temperature can be lowered with an increase in the strengths of either of the three- and four-body interaction terms, thereby confirming that a qubit attached to the common bath has to act as a cold qubit. This proves the presence of the common bath and consequently the few-body interaction terms extremely advantageous in scenarios where one is forced to operate in a parameter regime without a steady-state cooling. Our results indicate that the effect of an increase in the strength of the four-body interaction on lowering the steady-state temperature of the cold qubit is very small compared to the significant lowering of the cold-qubit temperature due to an increase in the strength of the three-body interaction term. We also demonstrate that our three-qubit two-bath setup can provide refrigeration for other types of qubit-bath interactions, for example, a reset interaction~\cite{linden2010,skrzypczyk2011,*brunner2012,*brunner2014,*brask2015}. We also comment on the possibility of obtaining a local cooling of one of the qubits by attaching the three-qubit system to a single reservoir, and discuss the thermodynamic consistency of such setups.

The paper is organized as follows. In Sec.~\ref{sec:model}, we describe the setup of the quantum absorption refrigerator constituted of three qubits and two baths, and discuss the few-body terms present in the qubit-bath interactions. We also discuss the quantum master equation and the derivation of the Lindblad superoperators. Section~\ref{sec:results} contains our main results on local steady-state cooling of a qubit in the cases of  two- and  single-bath refrigerators, and the effect of the few-body interaction terms on the temperature of the cold qubit in the steady state. The discussion on the use of the reset interaction between the qubits and the  baths for constructing the three-qubit two-bath quantum refrigerator is included in Sec.~\ref{sec:reset}. Section~\ref{sec:conclude} contains the concluding remarks.

\section{Model}
\label{sec:model}

We consider a system of three qubits, labeled by $1$, $2$, and $3$, and described by a Hamiltonian $H_S=H_0+gH_1$. Here $H_0$ is a local Hamiltonian given by  
\begin{eqnarray}
H_0=\sum_{i=1}^3\frac{E_i}{2}\sigma_i^z,
\label{eq:local-hamiltonian}
\end{eqnarray}
where $E_i$ is the energy gap between the levels of qubit $i$, such that the ground state $\ket{1}_i$ has an energy $-\frac{E_i}{2}$ while the excited state $\ket{0}_i$ has the energy $\frac{E_i}{2}$, and $\sigma_i^\alpha$ is the $\alpha$th component of the Pauli matrices, $\alpha=x,y,z$. On the other hand, $H_1$ represents the interaction among the three qubits, the strength of which is given by $g$. We are interested in an 
interaction of the form 
\begin{eqnarray}
H_1=\ket{010}\bra{101}+\text{h.c.}
\label{eq:int_Hamiltonian}
\end{eqnarray}
between the individual qubits in the three-qubit system. The motivation behind this choice lies in the fact that the dynamics of the system represented by $H_S=H_0+gH_1$ under Markovian qubit-bath interaction does not generate coherence in the system, resulting in diagonal reduced states of individual qubits obtained from the thermal state of the three-qubit system, once the system reaches its steady state. The interaction Hamiltonian $H_1$ is also in the same spirit as in the case of a quantum refrigerator constituted of three qubits and three thermal  baths~\cite{linden2010,correa2013,mitchison2015,das2019}, where a ``local" cooling of one of the qubits can be achieved by keeping each of the qubits in contact with a bath at a specific temperature, and by carefully choosing the qubit-bath interaction parameters.

\begin{figure}
\includegraphics[scale=0.5]{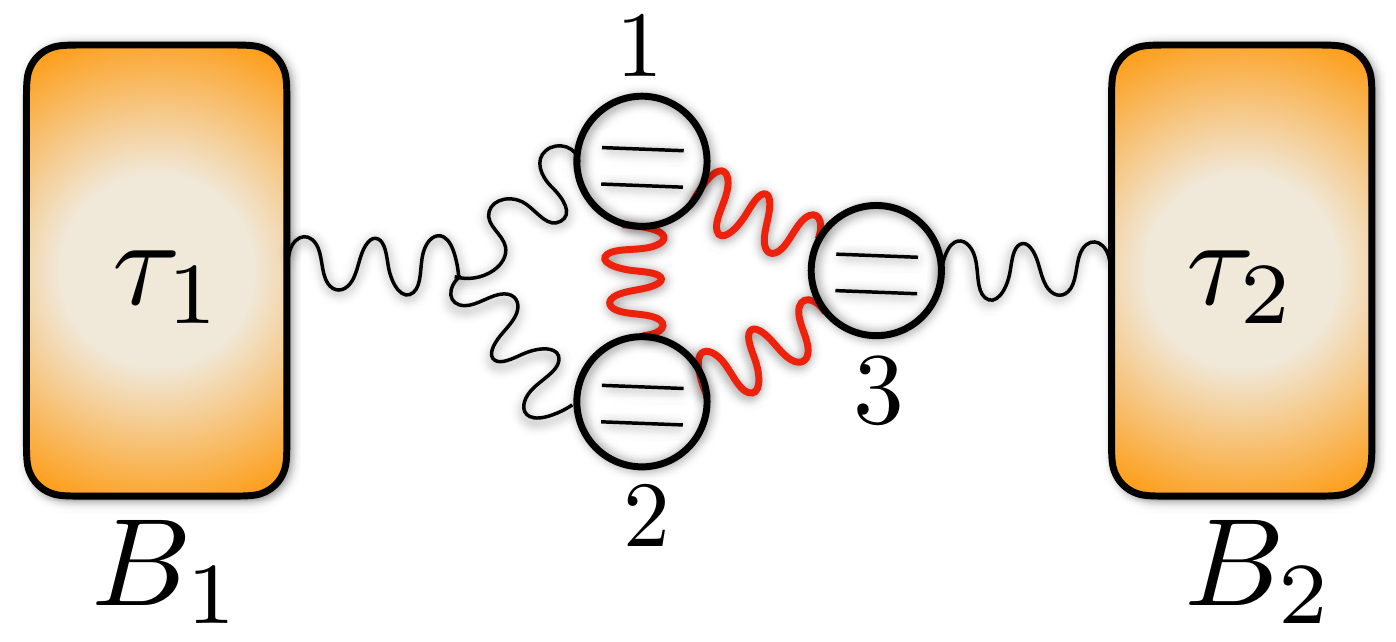}
\caption{\textbf{Schematic diagram representing the qubits-baths setup.} The qubits $1$ and $2$ are in contact with the bath $B_1$ at temperature $\tau_1$, while the qubit $3$ is connected to the bath $B_2$ at temperature $\tau_2$.}
\label{fig:schematic}
\end{figure}

We consider a scenario where at time $t=0$, qubits $1$ and $2$ are in thermal equilibrium with a common thermal heat bath, $B_1$, having a temperature $\tau_{1}$, while the qubit $3$ is in thermal equilibrium with the reservoir $B_2$, which is at temperature $\tau_2$ (see Fig.~\ref{fig:schematic}). We also assume that at $t=0$, the system Hamiltonian is solely represented by the local Hamiltonian $H_0$, i.e., $g=0$ at $t=0$, such that the state of the three-qubit system at $t=0$ is given by $\rho_S=\bigotimes_{i=1}^3\rho_i$. Here,  
\begin{eqnarray}
\rho_{i}= r_i\ket{0}_i\bra{0}+(1-r_i)\ket{1}_i\bra{1},
\label{eq:qubit-state}
\end{eqnarray}
with $r_i=Z_i^{-1}\exp (-E_i/2k_B T_i)$ being the probability that the qubit $i$ is in excited state, and $Z_i=\sum_{\alpha=1}^2\exp ((-1)^{\alpha}E_i/2k_B T_{i})$ is the partition function of the qubit $i$, $k_B$ being the Boltzmann constant, and $T_i$ being the temperature of the qubit $i$. At $t=0$, $T_1=T_2=\tau_{1},T_3=\tau_2$.

For each of the baths, we consider a canonical description of a Markovian thermal bosonic reservoir, constituted of an infinite collection of harmonic oscillators. In this paper, we assume that both of the bosonic baths are characterized by Ohmic spectral functions, given by $J_i(E)=\delta_{i} E\exp(-E/\Omega)$ corresponding to the bath $i$, where $\delta_i$ is the dimensionless qubit-bath interaction strength, and $\Omega$ is a cutoff frequency, identical for both of the baths, which leads to a bath memory time $\sim$ $\Omega^{-1}$.  We work in the Markovian regime, with a value of $\Omega$ large enough such that the bath memory time is much smaller than all other relevant time scales. The Hamiltonian for the bath $B_i$, $i=1,2$, is given by 
\begin{eqnarray}
H_{B_i}=\int_{0}^{\omega_i}\tilde{\omega}\eta_{i,\omega^\prime}^{\dagger}\eta_{i,\omega^\prime}d\omega^\prime,
\label{eq:bath-hamiltonian}
\end{eqnarray} 
where $\eta_{i,\omega^\prime}^{\dagger}(\eta_{i,\omega^\prime})$ is the bosonic creation (annihilation) operator corresponding to the  mode $\omega^\prime$ of the  bath $i$, having units of $\frac{1}{\sqrt{\omega^\prime}}$, and $\omega_i$ is the maximum frequency of the bath $i$. $\tilde{\omega}$ is a constant having the unit of frequency. In the situation described in Fig.~\ref{fig:schematic}, the full bath Hamiltonian is given by $H_B=\sum_{i=1}^2H_{B_i}$.

The system-bath coupling considered in the above set-up is given by  $H_I=H_I^{(1)}+H_I^{(2)}$, where $H_I^{(j)}$ denotes the interaction between the bath $j$, $j=1,2$, and the system. As depicted in Fig.~\ref{fig:schematic}, the bath $B_1$ interacts with qubits $1$ and $2$, while the bath $B_2$ interacts with the qubit $3$ only. The qubit-bath interaction Hamiltonian $H_I^{(2)}$, therefore, is given by 
\begin{eqnarray}
H_{I,2}^{(2)}&=&\int_{0}^{\omega_2}h_2(\omega)\sqrt{\tilde{\omega}}(\sigma_3^+\eta_{2,\omega}+\sigma_3^-\eta_{2,\omega}^{\dagger})d\omega,
\label{eq:single_qubit_bath_interaction}
\end{eqnarray}
where the ``2" in the subscript indicates that the interaction Hamiltonian is constituted of two-body interaction terms between the qubit $3$ and the bosonic modes of the bath $B_2$.  On the other hand, $H_I^{(1)}$ reads 
\begin{eqnarray}
H_I^{(1)}=H_{I,2}^{(1)}+\kappa_1 H_{I,3}^{(1)}+\kappa_2 H_{I,4}^{(1)}, 
\label{eq:int}
\end{eqnarray}
with 
\begin{eqnarray}
H_{I,2}^{(1)}&=&\sum_{i=1}^2\int_{0}^{\omega_{1}}h_1(\omega)\sqrt{\tilde{\omega}}(\sigma_i^+ \eta_{1,\omega}+\sigma_i^- \eta_{1,\omega}^{\dagger})d\omega,\nonumber\\
H_{I,3}^{(1)}&=&\int_{0}^{\omega_{1}}h_1(\omega)\tilde{\omega}(\sigma_1^+\sigma_2^-+\sigma_1^-\sigma_2^+)\eta_{1,\omega}^{\dagger}\eta_{1,\omega}d\omega,\nonumber \\
H_{I,4}^{(1)}&=&\int_{0}^{\omega_1}\int_{0}^{\omega_1}\tilde{\omega}\left[\sigma_1^+\sigma_2^+\eta_{1,\omega}\eta_{1,\omega^{\prime}}+\sigma_1^-\sigma_2^-\eta_{1,\omega}^{\dagger}\eta_{1,\omega^{\prime}}^{\dagger}\right]\nonumber\\ 
&&\times h_1(\omega)h_1(\omega^\prime)\delta(E_1+E_2-\omega-\omega^{\prime})d\omega d\omega^{\prime}.
\label{eq:common_qubit-bath-interaction}
\end{eqnarray}
Here, the Hamiltonians $H_{I,n}^{(2)}$, $n=2,3,4$, represent interactions between qubits $1$ and $2$ with the bosonic bath modes of bath $B_1$ via $n$-body interaction terms, $\kappa_1$ and $\kappa_2$ are the relative strengths of $H_{I,3}^{(1)}$ and $H_{I,4}^{(1)}$ with respect to $H_S+H_B+H_{I,2}^{(1)}+H_{I,2}^{(2)}$, $\sigma_i^{\pm}=\frac{1}{2}(\sigma^x_i\pm i\sigma^y_i)$ with $i=1,2,3$, and $\sqrt{\tilde{\omega}}h_i(\omega)=\sqrt{J_i(\omega)}$ with $i=1,2$. The presence of $\kappa_1$ and $\kappa_2$ gives us a handle to tune the strengths of three- and four-body interactions, relative to the other terms in the Hamiltonian. The Hamiltonian $H_{I,3}^{(1)}$ represents the interaction between the qubit pair $(1,2)$ and a common bosonic mode of the bath $B_1$ via the number operator, and is a natural consequence of having a single bath interacting with the two qubits
$1$ and $2$.  The full Hamiltonian representing the closed system constituted of the three-qubit system and the two baths can, therefore, be represented as 
\small 
\begin{eqnarray}
H&=&H_S+H_B+H_I,\nonumber \\
&=& (H_0+gH_1)+\sum_{i=1}^2H_{B_i}+\sum_{i=1}^2 H_{I,2}^{(i)}
+\kappa_1H_{I,3}^{(1)}+\kappa_2H_{I,4}^{(1)}.\nonumber\\ 
\label{eq:total_hamiltonian}
\end{eqnarray}\normalsize

We point out here that in the absence of the three- and four-body interaction terms (i.e., $\kappa_1=\kappa_2=0$), the common effect of the bath $B_1$ on the qubits $1$ and $2$ does not come into consideration, and the setup behaves qualitatively similar to a three-qubit three-bath construction of a quantum absorption refrigerator (for example~\cite{das2019}).

\noindent\textbf{Note:} The interaction Hamiltonian $H_{I,4}^{(1)}$ represents a four-body interaction between the qubit pair $(1,2)$ and two different bosonic modes of the bath $B_1$. There are different possibilities of four-body interactions between the qubits $1$, $2$, and the bath modes.  We choose one such possibility as the form of $H_{I,4}^{(1)}$ (Eq.~(\ref{eq:common_qubit-bath-interaction})) in order to check whether the presence of a four-body interaction term can affect the behavior of the model in a significant way. In principle, one can include all such four-body interaction terms obeying the energy conservation principle. 



\subsection{Quantum Master Equation and Lindblad Operators}
\label{subsec:qme}

If the interaction of strength $g$ among the three qubits is turned on for $t>0$, the system undergoes a dynamics according to the quantum master equation (QME) in the Lindblad form~\cite{breuer2002,rivas2012},
\begin{eqnarray}
\dot{\rho_S}=-\text{i}[H_S,\rho_S]+\mathcal{L}(\rho_S),
\label{eq:qme}
\end{eqnarray}
where $\mathcal{L}(.)$ is the non-unitary term determined according to the type of the thermal baths attached to the system, and the interactions of the thermal baths with the system. The general form of the Lindblad superoperator $\mathcal{L}(.)$ is of the form\small 
\begin{eqnarray}
\mathcal{L}(\varrho)=\sum_{i=1}^{3}\sum_{j}\gamma_{i,j} \left[L_{i,j}\varrho L_{i,j}^\dagger-\frac{1}{2}\{L_{i,j}^\dagger L_{i,j},\varrho\}\right],
\label{eq:lindblad_superoperator}
\end{eqnarray}\normalsize
where $\gamma_{i,j}$ is a rate which may or may not depend on the energy of the system, and $L_{i,j}$ is a Lindblad operator~\cite{breuer2002,rivas2012} corresponding to a process for qubit $i$ due to its reservoir and occurring at the rate $\gamma_{i,j}$, $j$ being a process index.  Here and in the rest of the paper, we consider the dimensionless time $t$ in units of $K\tilde{t}/\hbar$, the dimensionless energies $\{E_i\}$, and $g$ in units of $K$, and the dimensionless temperatures $\tau_1$, $\tau_2$, and $\{T_i\}$ to be in units of $k_B\tilde{T}_i/K$, where $\tilde{t}$ and $\tilde{T}_i$ are the actual time and the absolute temperature, and $K$ is an arbitrary constant having the dimension of energy determined according to the experimental setup used to implement the total Hamiltonian $H$.  We also assume Markovian system-bath interaction.

The form of the Lindblad operators can be derived by going over to the interaction picture generated by $H_S+H_B$, and successively applying Born 
Markov, 
and rotating-wave approximations~\cite{breuer2002,rivas2012}. In the situation where $g \sim E_i$, the rotating-wave approximation is valid if and only if the typical energy differences of the system $\max\{\gamma_{i}\} \ll \min\{E_i,g\}$~\cite{breuer2002,rivas2012,correa2013,mitchison2015}. In this scenario, the nonunitary term in Eq. (\ref{eq:qme}) can be written as
\begin{eqnarray}
\mathcal{L}(\rho_S)&=&\sum_{j=1}^2\sum_{n=2}^4 \sum_{\{E\}}\gamma_{j,n}(E)\Big[L_{j,n}^E\rho_S{L_{j,n}^{E}}^\dagger\nonumber\\
&&-\frac{1}{2}\left\{{L_{j,n}^E}^\dagger L_{j,n}^E,\rho_S\right\}\Big],
\label{eq:strong:dynamical}
\end{eqnarray}
where $j$ denotes the bath index, and $\{L_{j,n}^E\}$ is the set of Lindblad operators corresponding to the transitions due to the bath $B_i$ between the eigenstates of $H_{S}$ having energy-gap $E$, where the transitions happen at the rates $\gamma_{j,n}(E)$. Determination of $L_{j,n}^E$ is equivalent to decomposing the system operators from the system-bath interaction terms $H_{I,n}^{(j)}$, $j=1,2$, $n=2,3,4$, in $H_I$ into eigenoperators $L_{j,n}^E$ corresponding to energy $E$ of $H_{S}$. 
All the Lindblad operators are non-Hermitian,  satisfying $[H_S,L_{j,n}^{E}] = -E L_{j,n}^E$, and are defined as
\begin{eqnarray}
L_{1,2}^{E} &=& \sum_{i=1}^{2}\sum_{E_{a^\prime}-E_{a}=E}\ket{a}\bra{a}(\sigma_i^+ +\sigma_i^-)\ket{a^\prime}\bra{a^\prime} ,\nonumber\\
L_{1,3}^{E} &=& \sum_{E_{a^\prime}-E_{a}=E}\ket{a}\bra{a}(\sigma_1^+\sigma^-_2+\sigma_1^-\sigma^+_2)\ket{a^\prime}\bra{a^\prime},\nonumber\\
L_{1,4}^{E} &=& \sum_{E_{a^\prime}-E_{a}=E}\ket{a}\bra{a}(\sigma_1^+\sigma^+_2+\sigma_1^-\sigma^-_2)\ket{a^\prime}\bra{a^\prime},\nonumber\\
L_{2,2}^E &=& \sum_{E_{a^\prime}-E_{a}=E}\ket{a}\bra{a}(\sigma_3^++\sigma^-_3)\ket{a^\prime}\bra{a^\prime},  
\label{eq:lindblad_general}
\end{eqnarray}
where $\ket{a}$, $\ket{a^\prime}$ are non-degenerate eigenstates of $H_S$ with energies $E_{a}$, $E_{a^\prime}$, respectively, and $L_{j,n}^{-E}={L_{j,n}^{E}}^\dagger$. The explicit forms of the Lindblad superoperators are obtained from these expressions by using the eigenvectors $\{\ket{\mathbf{0}},\ket{\mathbf{1}},\cdots,\ket{\mathbf{7}}\}$ of $H_S$ given by
\begin{eqnarray}
\label{equ:eigenstates}
\ket{\mathbf{0}}&=& \ket{000}, \ket{\mathbf{1}} = \ket{001}, \ket{\mathbf{2}} = \ket{100},\nonumber \\ 
\ket{\mathbf{3}}&=& \ket{011}, \ket{\mathbf{4}} = \ket{110}, \ket{\mathbf{5}} = \ket{111},\nonumber \\ 
\ket{\mathbf{6}}&=& (\varepsilon_+\ket{010}+\ket{101})/\sqrt{1+|\varepsilon_+|^2},\nonumber \\ 
\ket{\mathbf{7}}&=& (\varepsilon_-\ket{010}+\ket{101})/\sqrt{1+|\varepsilon_-|^2},
\label{eq:all_states}
\end{eqnarray}
where
\begin{eqnarray}
\varepsilon_\pm&=&\frac{1}{2g}\left[E_3-\Delta E\pm\tilde{E}\right],
\end{eqnarray}
with $\Delta E=E_2-E_1$ and $\tilde{E}=[(E_3-\Delta E)^2+4g^2]^{\frac{1}{2}}$, while the  eigenenergies corresponding to these states are 
\begin{eqnarray}
\label{equ:eigenenergies}
E_{\mathbf{0}}&=&\frac{1}{2}(E_{12}+E_3),
E_{\mathbf{1}}=\frac{1}{2}(E_{12}-E_3),\nonumber\\  E_{\mathbf{2}}&=&\frac{1}{2}(E_3+\Delta E), E_{\mathbf{3}}=-\frac{1}{2}(E_3+\Delta E),\nonumber \\ E_{\mathbf{4}}&=&-\frac{1}{2}(E_{12}-E_3), E_{\mathbf{5}}=-\frac{1}{2}(E_{12}+E_3),\nonumber\\  E_{\mathbf{6}}&=&\tilde{E}/2, E_{\mathbf{7}}=-\tilde{E}/2,
\end{eqnarray}
with $E_{12}=E_1+E_2$.
\section{Performance as a refrigerator}
\label{sec:results}

We now discuss the performance of the three-qubit system described by the Hamiltonian $H_S$ interacting with two reservoirs (see Sec.~\ref{sec:model}) as a quantum refrigerator, where the local temperature of one of the qubits is lowered as an effect of the dynamics of the system. Note that the availability of the two baths $B_1$ and $B_2$ of temperatures $\tau_1$ and $\tau_2$ ($\tau_1<\tau_2$), respectively, implies that the temperature of a chosen qubit can always be lowered to $\tau_1$ by keeping it in equilibrium with the bath $B_1$. Therefore, operation of the machine as a refrigerator for one of the qubits is beneficial if and only if the temperature of the qubit can be lowered below $\tau_1$ during the dynamics. In the subsequent discussions, we shall demonstrate that this is indeed the case for the three-qubit two-bath setup discussed in Sec.~\ref{sec:model}.  

\subsection{Working principle}
\label{subsec:working_principle}

At $t=0$, the three-qubit system is in contact with the thermal reservoirs $B_1$ and $B_2$ as shown in Fig.~\ref{fig:schematic}. 
Let us denote the probabilities of finding the system in the energy eigenstates $\ket{010}$ and $\ket{101}$, corresponding to the energy eigenvalues $E_{010}=(E_3-\Delta E)/2$ and $E_{101}=(\Delta E-E_3)/2$, by $p_{010}$ and $p_{101}$, respectively. When the interaction term $H_1$ in the system Hamiltonian $H_S$ is turned on, it incorporates a transition $\ket{010}\leftrightarrow\ket{101}$ between the states $\ket{010}$ and $\ket{101}$. We begin by assuming the weak-coupling regime of operation, given by $g \ll \min\{E_i\}$, 
ensuring that the energy eigenvalues and eigenstates of $H_S$ are not drastically modified once the interaction Hamiltonian $H_1$ is turned on (for $g\rightarrow 0$, $E_{\mathbf{6}}\rightarrow E_{010}$ and $E_{\mathbf{7}}\rightarrow E_{101}$). 
We also assume the energy-preserving interactions between the qubits; i.e., we require that $H_1$  commutes with $H_0$, which requires that 
$\Delta E=E_3$, i.e., $E_2=E_3+E_1$. In this scenario, if the probability of the transition $\ket{010}\rightarrow\ket{101}$ can be increased (i.e., $p_{101}>p_{010}$), it may lead to a cooling of qubit $1$ in the sense that the population of the ground state $\ket{1}$ of qubit $1$ increases due to a transfer of excitation from qubit $1$ to either qubit $2$, or qubit $3$, or both. The qubit among $2$ and $3$ receiving the transferred energy undergoes a heating by an increase in the population of its excited state.

For $E_{101}=E_{010}=0$, one can write $p_{010}$ and $p_{101}$ explicitly in terms of the energies of the individual qubits and their equilibrium temperatures as 
\begin{eqnarray}
p_{010} &=& \prod_{i=1}^3 Z_i^{-1}\exp\left[\frac{\Delta E}{2k_B}\left\{\frac{1}{\tau_1}-\frac{1}{\tau_2}\right\}\right],\nonumber\\ 
p_{101} &=& \prod_{i=1}^3 Z_i^{-1}\exp\left[-\frac{\Delta E}{2k_B}\left\{\frac{1}{\tau_1}-\frac{1}{\tau_2}\right\}\right], 
\label{eq:prob}
\end{eqnarray}
where $Z_i$ is the partition function for qubit $i$ (see Eq.~(\ref{eq:qubit-state}) and subsequent discussions). It is clear from Eq.~(\ref{eq:prob}) that $p_{010}=p_{101}$ for $\tau_1=\tau_2$, thereby prohibiting any transition between the energy levels $E_{101}$ and $E_{010}$. Let us assume that $\tau_1<\tau_2$, so that $\frac{1}{\tau_1}-\frac{1}{\tau_2}>0$. Under this condition, we note the following important points.   
\begin{enumerate}
    \item For $\Delta E=0$, $p_{010}=p_{101}$, forbidding a transition between the energy levels $E_{010}$ and $E_{101}$. Therefore, no cooling of either of the qubits $1$ or $2$ takes place, which is consistent with the symmetry of the setup. 
    \item For $\Delta E>0$ ($\Delta E<0$), $p_{101}>p_{010}$ ($p_{010}>p_{101}$), implying a cooling of qubit $1$ (qubit $2$). 
    \item Note that a cooling (heating) of qubit $3$ takes place for $\Delta E>0$ ($<0$). Therefore, $\Delta E>0$ provides a situation where two out of the three qubits may undergo cooling when the interaction Hamiltonian is turned on. 
\end{enumerate}
On the other hand, for $\tau_1>\tau_2$, $\frac{1}{\tau_1}-\frac{1}{\tau_2}<0$, implying $p_{010}>p_{101}$ ($p_{010}<p_{101}$) for $\Delta E>0$ ($<0$), and the algorithmic cooling and heating of the different qubits occur accordingly. 


We point out here that although the above consideration of algorithmic cooling or heating of individual qubits indicates the trend of the temperature of the qubits when the dynamics is started by turning on $g$, it does not predict whether an overall cooling or heating of the chosen qubit will occur once the system reaches its steady state. (Similarly, it is difficult to predict - without following the actual evolution - the possibility of cooling in the strong-coupling regime.) The steady state of the system is a result of the initial state of the system, the system Hamiltonian, the types of the bath, as well as the specific form of the qubit-bath interactions. Note here that a three-qubit quantum refrigerator can also be constructed using three thermal baths instead of two (for example, see~\cite{linden2010,correa2013,mitchison2015}). A steady-state cooling of qubit $1$ occurs under the weak- as well as strong-coupling condition of the three-qubit three-bath quantum refrigerator~\cite{correa2013,mitchison2015} using bosonic thermal baths. We shall discuss the salient features of these machines, and the specific differences between their properties and the properties of the three-qubit two-bath refrigerator model in subsequent sections.

\subsection{Dynamics of cold-qubit temperature}
\label{subsec:dynamics_Bosonic}

\begin{figure*}
\includegraphics[width=\textwidth]{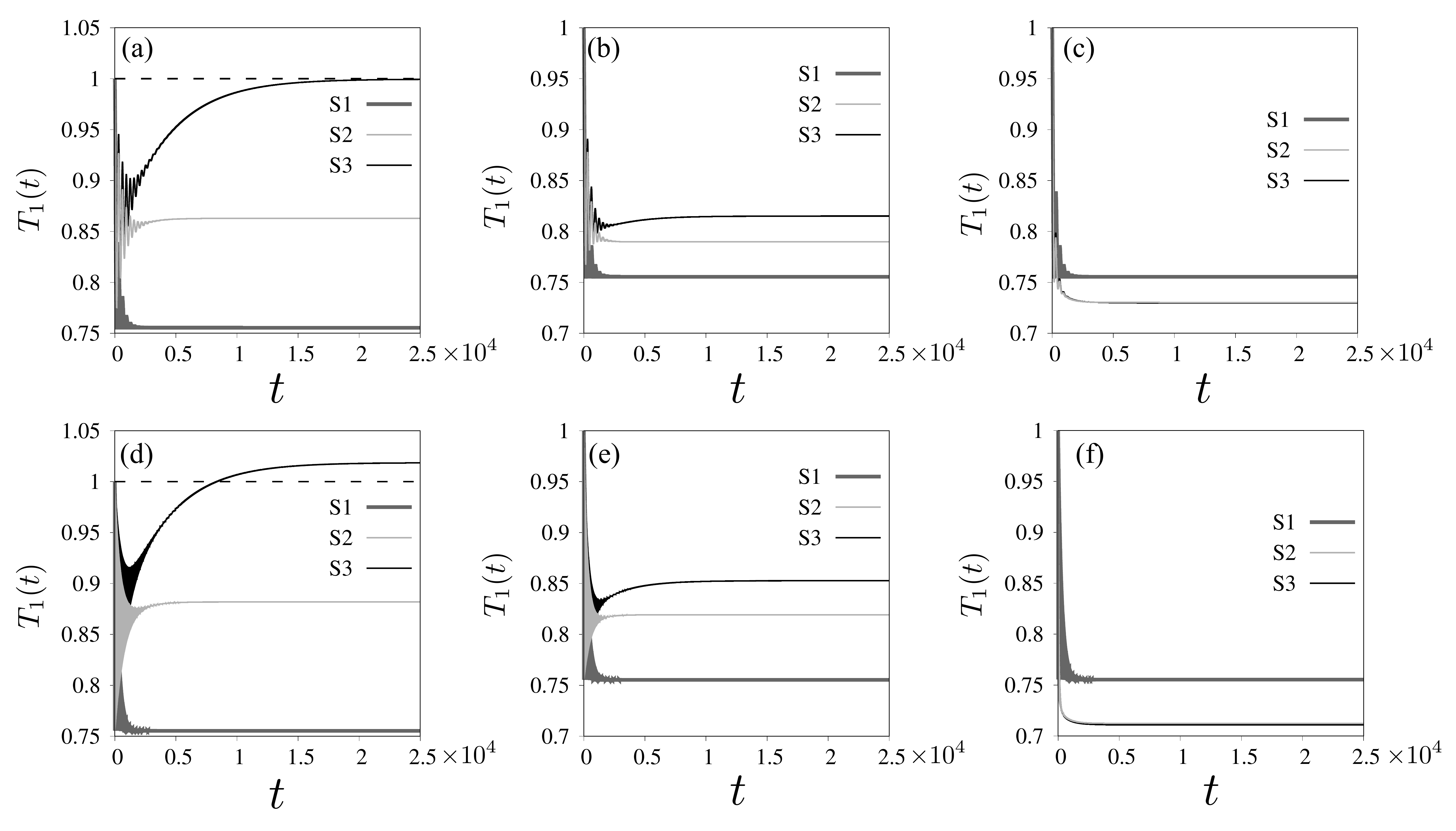}
\caption{
\textbf{Dynamics of cold-qubit temperature in the weak- and strong-coupling operating regimes.} We choose $g=10^{-2}$  (``weak-coupling'') in panels (a), (b), and (c), and 
$g=0.5$ (``strong-coupling'') in panels (d), (e), and (f). The values of \(\kappa_1\) differ in the different panels, with \(\kappa_1=\kappa_2=0\) for panels $(a)$ and $(d)$, \(\kappa_1=1,\kappa_2=0\) for $(b)$ and $(e)$, \(\kappa_1=2,\kappa_2=0\) for $(c)$, and \(\kappa_1=4,\kappa_2=0\) for $(f)$. \(\kappa_2\) is kept at zero throughout. 
The definitions of the types of dynamics denoted by \textbf{S1-S3} are as given in Sec.~\ref{subsubsec:two-body}. For all of the panels (a)-(f), the values of the qubit-bath interaction parameters are chosen to be $i)\;\;\delta_1=10^{-8},\delta_2=10^{-4}$ for \textbf{S1}, $ii)\;\;\delta_1=10^{-4},\delta_2=10^{-5}$ for \textbf{S2}, and $iii)\;\;\delta_1=10^{-4},\delta_2=10^{-8}$ for \textbf{S3}. The horizontal black dashed lines in panels $(a)$ and $(d)$ correspond to the cold bath temperature $\tau_1$. For the other panels, the same  coincide with  
the horizontal upper boundaries. All quantities plotted are dimensionless.} 
\label{fig:harmonic}
\end{figure*} 

Once the interaction of strength $g$ among the three qubits is turned on for $t>0$, the system is out of equilibrium and undergoes a dynamics according to Eq.~(\ref{eq:strong:dynamical}). A cooling of the qubit, say, $1$, occurs at a finite time instant $t$ during the evolution if the temperature $T_1(t)<\tau_1$. Since the Markovian qubit-bath interactions do not generate coherence in the system, a \emph{local} temperature of qubit $1$ can be defined using the fact that the local density matrices of the individual qubits are always diagonal during the evolution of the system as well as in the  steady state. Considering the diagonal form given in Eq.~(\ref{eq:qubit-state}) for the state $\rho_1(t)$ of qubit $1$ during the dynamics, its temperature, as a function of time, is given by 
\begin{eqnarray}
T_1(t)=E_1\left[\ln\left(\frac{1-r_1(t)}{r_1(t)}\right)\right]^{-1},
\label{eq:cold_qubit_temperature}
\end{eqnarray}
where $\rho_1(t)=\text{Tr}_{23}[\rho_S(t)]$, $\rho_S(t)$ being the three-qubit state as a function of time according to the evolution of the system. Note that the local temperatures of qubits $2$ and $3$ can also be defined using equations similar to Eq.~(\ref{eq:cold_qubit_temperature}). 

We fix $\Delta E=E_3>0$, and focus on the dynamics of the temperature of qubit $1$, which we refer to as the \textit{cold qubit} from here onward. We solve the QME numerically using the Runge-Kutta 4th-order method, and determine the quantum state of the system as a function of the system as well as system-bath interaction parameters, and time. The temperature of the cold qubit, $T_1(t)$, as a function of time can be derived using Eq.~(\ref{eq:cold_qubit_temperature}). We consider the steady state of the dynamics of the cold qubit temperature to be achieved when 
\begin{eqnarray}
|T_{1}(t)-T_1(t-\delta t)|<10^{-9}
\label{eq:steady_state_criteria}
\end{eqnarray}
for a finite time interval $\Delta t\gg\delta t$. We have chosen $\Delta t=5\times 10^2$, and $\delta t=10^{-1}$ for all our computations reported in this paper. Let the criterion defined in Eq.~(\ref{eq:steady_state_criteria}) be satisfied for the first time during the dynamics at the time $t^\prime$. Then the system is considered to have achieved steady state at time $t_s=t^\prime -\Delta t$, and the steady-state temperature of the cold qubit is given by $T_1^s=T_1(t=t_s)$. We consider a \emph{steady-state cooling} (SSC) to be the scenario where the dynamics of the system leads to a steady-state temperature, $T_1^s$, of the cold qubit which is less than its initial temperature, $\tau_1$. In this situation, the qubit $1$ represents the object to be cooled, and the rest of the system and the reservoirs construct the refrigerator that dynamically takes the object to a temperature lower than its initial temperature, and keeps it at the lowered temperature as $t\rightarrow\infty$. On the other hand, a \emph{transient cooling} (TC) of the cold qubit takes place if $T_1(t)<\tau_1$ at a time $t<t_s$. Such transient cooling can happen for $t \ll t_s$.

\subsubsection{Refrigeration with \texorpdfstring{$\kappa_1=\kappa_2=0$}{k1=k2=0}}
\label{subsubsec:two-body}
We first consider a situation where the interaction between the qubits $1$, $2$ and the bath $B_1$ is determined only by the two-body interaction term $H_{I,2}^{(1)}$ ($\kappa_1=\kappa_2=0$ in Eq.~(\ref{eq:total_hamiltonian})).
The temperature of the cold qubit may exhibit a variety of dynamics. However, subject to cooling occurring in  qubit $1$, its dynamic profiles can, in general, be categorized into three types, namely, \textbf{S1}; when SSC is better  than TC; \textbf{S2}, when TC is better than SSC, and \textbf{S3.} when TC takes place without an SSC. Let us assume that $T_1^t$ is the minimum temperature achieved by the cold qubit during transient dynamics for $t<t_s$. Using $T_1^t$ and $T_1^s$, the three scenarios for cooling of qubit $1$ can be described as follows: 
\begin{itemize} 
\item[\textbf{S1.}]\textbf{SSC better than TC.} This situation can be characterized by $T_1^s<T_1^t <\tau_1$, and assuming continuity, \(T_1^s=T_1^t <\tau_1\).   

\item[\textbf{S2.}]\textbf{TC better than SSC.} This refers to a cold qubit dynamics with $T_1^t<T_1^s<\tau_1$. 

\item[\textbf{S3.}]\textbf{TC without SSC.} In the third scenario, only transient cooling occurs with no or negligible SSC, 
i.e.,
\(T_1^t<\tau_1 \lesssim T_1^s\). 
\end{itemize}
We demonstrate these three types of temperature dynamics for the qubit $1$ in panels (a) and (d) of Fig. \ref{fig:harmonic},
where we keep $\kappa_1$ and $\kappa_2$ at zero values. The values of the energies of the individual qubits, the temperatures of the baths, and the cut-off frequency are fixed at $E_1=1,E_2=2,E_3=1$, $\tau_1=1,\tau_2=5$, and $\Omega=10^3$ for demonstration. Unless otherwise mentioned, we keep these values fixed throughout the paper.  The values of the rate constants, $\gamma_{1,2}(E)$ and $\gamma_{2,2}(E)$, are determined from the energies of the individual qubits as well as the temperatures of the thermal reservoirs (see Appendix~\ref{app:specific_lindblad}). For demonstration, we choose two different values of the qubit-qubit interaction strength $g$, representing the weak- $\left(g=10^{-2},\;\max\{\gamma_i\} \ll g \ll \min\{E_i\}\right)$ and the strong-coupling $\left(g=0.5,\;\max\{\gamma_i\} \ll g\approx E_i\right)$ operating regimes of the  machine, such that the rotating wave approximation $\left(\max\{\gamma_i\} \ll \min\{g,E_i\}\right)$ remains valid  in both cases. Unless otherwise mentioned, we use these two values of $g$ as representative of the two regimes in all our demonstrations in this paper. The oscillations in the transient dynamics of the cold-qubit temperature increase with an increase in the value of $g$ (see panels (a) and (d) of Fig.~\ref{fig:harmonic}), implying that a steady-state cooling can, in principle, be achieved without an oscillatory dynamics in the strong-coupling regime when $g$ is high enough (cf.~\cite{mitchison2015} for the similar dynamics in the case of a three-qubit three-bath refrigerator). 

Note that the different types of cooling of qubit $1$ reported above are qualitatively similar to the dynamics of cold-qubit temperature obtained in~\cite{das2019}, where a three-qubit three-bath setup of the refrigerator was used. As already mentioned in Sec.~\ref{sec:model}, the reason behind such similarity is the absence of the common effect of the bath $B_1$ due to the absence of the few-body interaction terms in the system-bath interaction. Note also that as the qubit pair, $1$ and 2, is interacting with a common bath, it is difficult to conceptualize 
and 
determine the direction of heat flow from or to either of these qubits, and to explain the cooling process of the individual qubits accordingly. Nevertheless, we observe that a \textit{local cooling} occurs in the qubit $1$ during the dynamics of the three-qubit system -- its indication being a reduction in the value of $r_1$ over time. Therefore, in the subsequent sections of this paper, we will refer the cooling of qubit $1$ as ``local" cooling.  

\begin{figure*}
\includegraphics[width=0.7\textwidth]{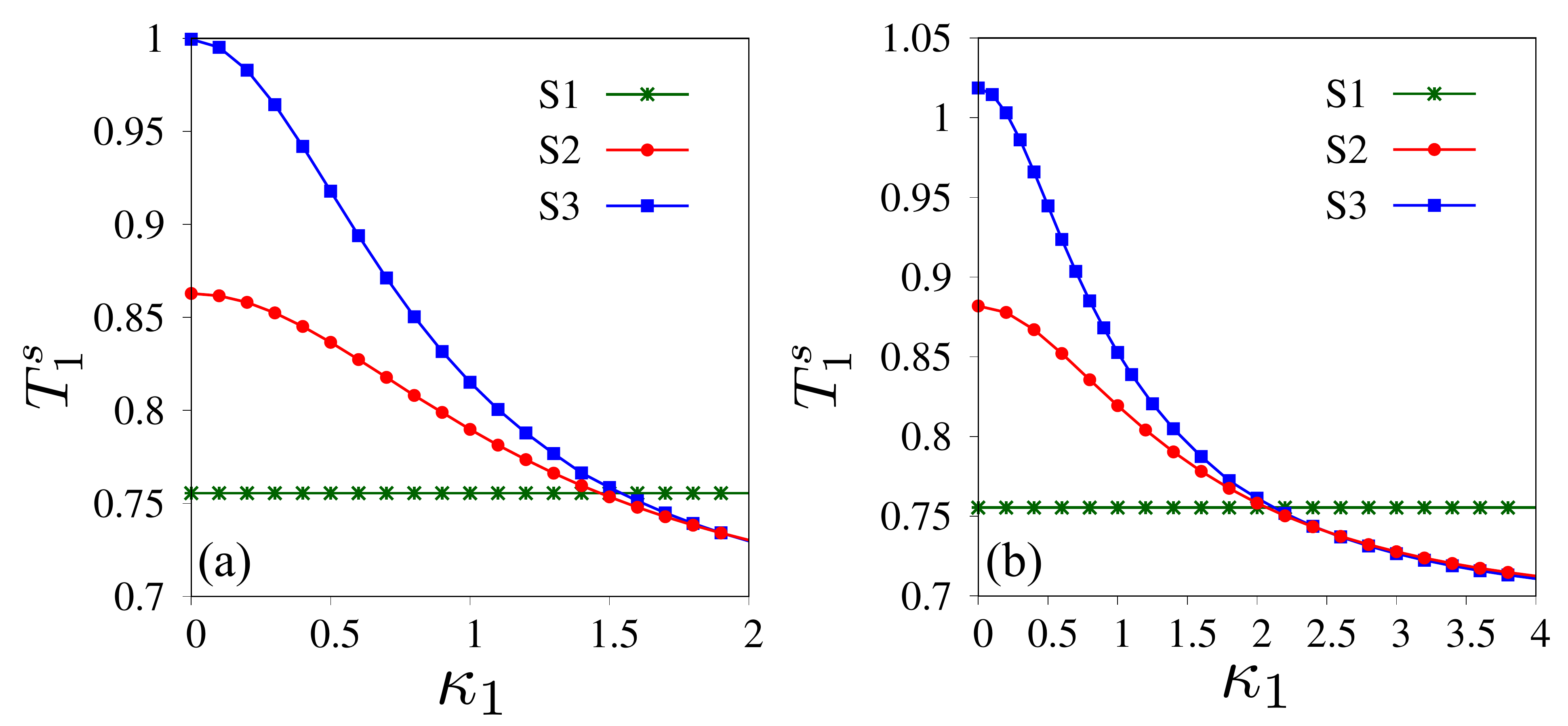}
\caption{\textbf{Behavior of steady-state temperature with three-body interaction.} Variation of $T_1^s$ as a function of $\kappa_1$, with $\kappa_2=0$ in the (a) weak- ($g=10^{-2}$) and (b) strong-coupling ($g=0.5$) regimes. The values of the qubit-bath interaction parameters are the same as in Fig. \ref{fig:harmonic}. All quantities plotted are dimensionless.}
\label{fig:kappa1}
\end{figure*}
\begin{figure*}
\includegraphics[width=\textwidth]{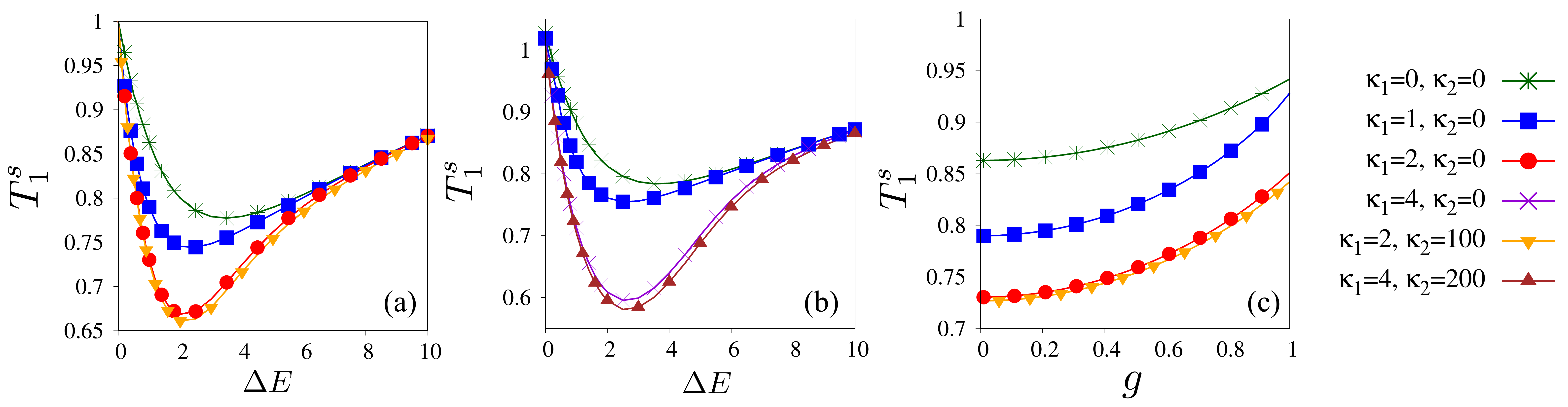}
\caption{ \textbf{Variations of the steady-state temperature of the cold qubit as function of system parameters.} Behaviours of $T_1^s$  with varying $\Delta E$  are depicted in  (a) (weak-coupling: $g=10^{-2}$) and (b) (strong-coupling: $g=0.5$), while (c) exhibits the variation of $T_1^s$ with $g$. The values of the qubit-bath interaction parameters are $\delta_1=10^{-4},\delta_2=10^{-5}$. Note that the choice of the $(\kappa_1,\kappa_2)$ pair dictates the relative strengths between the single-body term, and two-body, three-body, and four-body interactions. All quantities plotted are dimensionless.}
\label{fig:calibration}
\end{figure*}
   
\subsubsection{Better local cooling in the presence of three-body interaction}
\label{subsubsection:three-body}
We now investigate the effect of the three-body interaction term in the interaction between the qubits $1$, $2$, and the bath $B_1$, which is crucial in incorporating the effect of the simultaneous interaction of a common bath mode with both the qubits $1$ and $2$. Note that for $0\leq\kappa_1< 1$ and $\kappa_2=0$, the effect of $H_S+H_B+(H_{I,2}^{(1)}+H_{I,2}^{(2)})$ in the total Hamiltonian  (see Eq.~(\ref{eq:total_hamiltonian})) dominates, while for $\kappa_1> 1$, the leading term in the total Hamiltonian is $H_{I,3}^{(1)}$. However, note that a nonzero value of $\kappa_1$ also leads to non-zero transition rates $\gamma_{1,3}(E)$ (see Appendix~\ref{app:specific_lindblad}), and contributes to it as $\kappa_1^2$. In order to preserve the validity of the rotating-wave approximation, $\kappa_1$ can only be increased up to a maximum possible value $\kappa_1^m$. In the parameter space of the three-qubit quantum refrigerator model constituted of two baths, $\kappa_1^m$ is a function of $\{E_i;i=1,2,3\}$, $g$, $\{\tau_i;i=1,2\}$, $\{\delta_i;i=1,2\}$, and $\Omega$ (see Eq.~(\ref{eq:kappa1})).
As the rotating-wave approximation sets  upper bounds on the values of the \(\gamma\), upper bounds are set on the \(\kappa_1\) and \(\kappa_2\).  The expressions of \(\kappa_1\) and \(\kappa_2\) are given in Eqs.~(\ref{eq:kappa1}) and (\ref{eq:kappa2}), respectively. For all our demonstrations, we choose $\kappa_1\leq \kappa_1^m$, so that $\gamma_{1,3}(E)$  $\sim 10^{-3}$ for the weak- and $\sim10^{-2}$ for the strong-coupling regimes, ensuring that $\gamma_{1,3}(E)$ has a sufficiently low value compared to the corresponding values of $g$, and therefore  satisfying the rotating-wave approximation. An increase in the value of  $\kappa_1$ results in $\gamma_{1,3}(E)$ being comparable to $g$, thereby violating the rotating-wave approximation.  

\begin{figure*}
\includegraphics[width=0.7\textwidth]{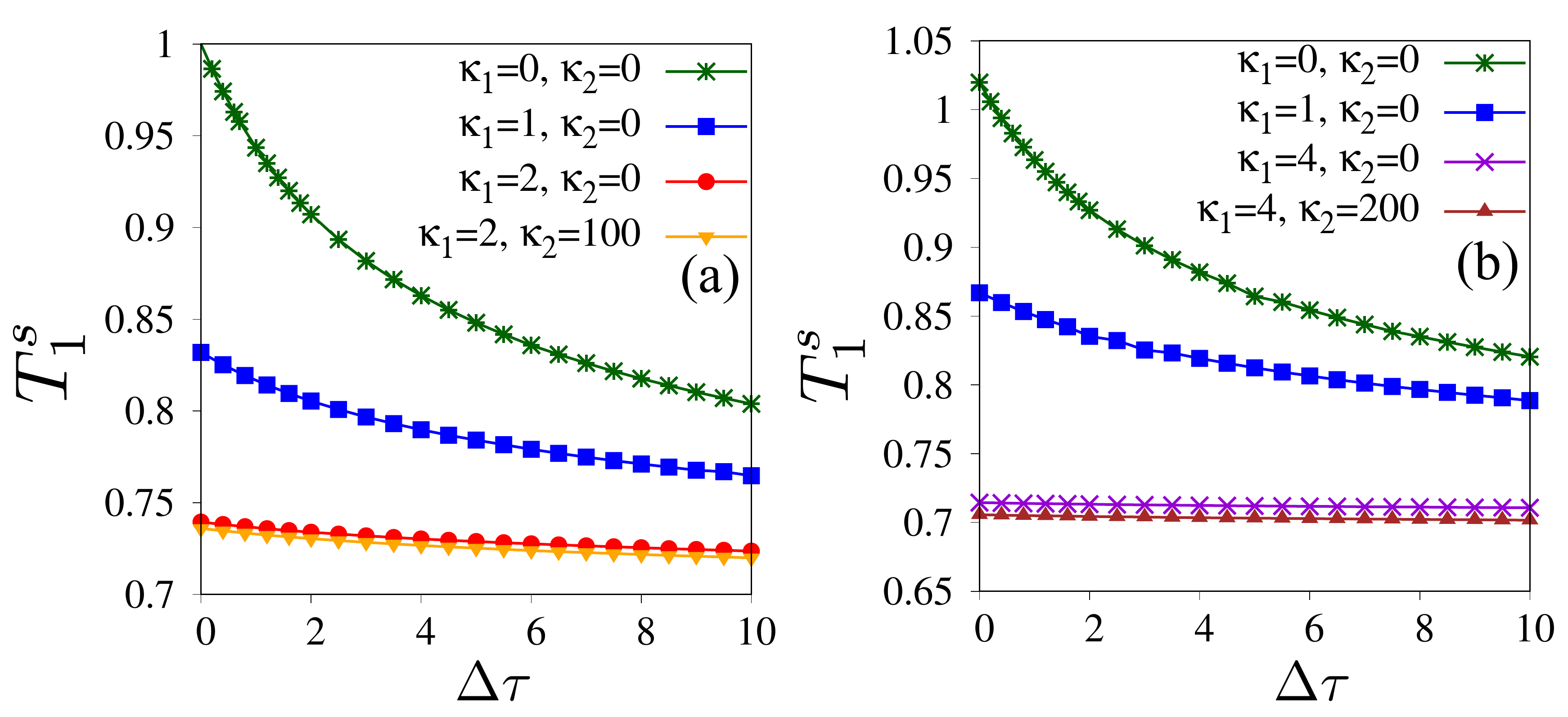}
\caption{\textbf{Variation of steady-state temperature of the cold qubit with respect to difference of temperatures of the two baths.} Variations of $T_1^s$ against $\Delta \tau$ in the (a) weak- ($g=10^{-2}$) and (b) strong- ($g=0.5$) coupling regimes are plotted here. The values of the qubit-bath interaction parameters are $\delta_1=10^{-4},\delta_2=10^{-5}$. All quantities plotted are dimensionless.}
\label{fig:temp}
\end{figure*}

In Figs.~\ref{fig:harmonic}(b)-(c) and Figs.~\ref{fig:harmonic}(e)-(f), we depict the effect of increasing $\kappa_1$ on the dynamics of the cold-qubit temperature for the weak- and strong-coupling regimes, respectively, while keeping the strength of the four-body interaction term zero ($\kappa_2=0$). It is clear from the figures that the steady-state temperature of the cold qubit  is significantly lowered from its value at $\kappa_1=0,\kappa_2=0$ for a non-zero strength of $H_{I,3}^{(1)}$. This effect of the three-body interaction term is  advantageous particularly when one is forced to work in the parameter regime where the dynamics is of \textbf{S2} or \textbf{S3} type for $\kappa_1=0$, $\kappa_2=0$, although a local steady-state cooling of qubit $1$ is desired. The effect of a non-zero $\kappa_1$ is more prominent in the strong-coupling regime of operation of the three-qubit quantum refrigerator with two baths. This is clearly seen from Figs.~\ref{fig:kappa1}(a)-(b), where we plot the variation of the steady-state temperature of the cold qubit as a function of $\kappa_1$ in the (a) weak- and (b) strong-coupling operating region, keeping $\kappa_2=0$. Note that 
\begin{enumerate}
\item[(i)] the effect of the three-body interaction on the steady-state temperature of the type-\textbf{S1} dynamics is negligible, while considerable lowering of $T_1^s$ takes place with increasing $\kappa_1$ for type-\textbf{S2} and type-\textbf{S3} dynamics; and 
\item[(ii)] when the three-body interaction term is dominating in the total Hamiltonian and the four-body interaction term is absent (i.e., $\kappa_1 >1$, $\kappa_2=0$), the steady-state temperature of qubit $1$ corresponding to both type-\textbf{S2} and type-\textbf{S3} dynamics tend to have identical behavior qualitatively as well as quantitatively with increasing $\kappa_1$. 
\end{enumerate}

At this point, it is logical to ask how $T_1^s$ varies with the system parameters and the temperatures of the baths, when the qubit-bath interaction corresponding to the bath $1$ has only two-body ($\kappa_1=\kappa_2=0$), and two-body as well as three-body ($\kappa_1>0,\kappa_2=0$) interactions. As discussed in Sec.~\ref{subsec:working_principle}, the temperature of qubit $1$ does not decrease unless $\Delta E>0$. However, in the absence of the three-body interaction term (i.e., for $\kappa_1=0$), $T_1^s$ has a non-monotonic variation with increasing $\Delta E$ in both weak- as well as strong-coupling regimes of the refrigerator. The variation increases in both regimes as the strength of the three-body interaction term increases [see Figs.~\ref{fig:calibration}(a) and~\ref{fig:calibration}(b)]. With increasing $\Delta E$, the values of $T_1^s$ decreases first, attains a minimum, and then increases again to achieve an almost constant value close to $\tau_1$. This implies that an optimum value of $\Delta E$ exists for a specific value of $\kappa_1$ in order to obtain the maximum possible local steady-state cooling. Increasing $\Delta E$ beyond the optimum value for a specific value of $\kappa_1$ would decrease the performance of the setup as a refrigerator for qubit $1$, and eventually lead to a situation where almost no local steady-state cooling will be obtained. The minimum value of $T_1^s$ decreases with increasing $\kappa_1$ while keeping $\kappa_2=0$, as depicted clearly in the figure.

In Fig.~\ref{fig:calibration}(c), we plot $T_1^s$ as a function of $g$, where the value  of $g$ changes from weak-coupling regime to strong-coupling regime, and the values of the parameters $\delta_i$ are chosen such that the rotating-wave approximation remains valid in both regimes. We investigate both situations where only two-body ($\kappa_1=\kappa_2=0$), and two- as well as three-body $(\kappa_1\neq 0, \kappa_2=0)$ interactions are present in $H_I^{(1)}$, but the four-body interaction term is absent ($\kappa_2=0$). As the figure demonstrates, $T_1^s$ increases monotonically with $g$, indicating a better local steady-state cooling being favored when the refrigerator is operating with a lower value of $g$.  With increasing strength of $\kappa_1$ keeping $\kappa_2=0$, $T_1^s$ decreases for a fixed value of $g$.     

We also investigate how the steady-state temperature of qubit $1$ varies with the temperatures of the baths $B_1$ and $B_2$. We find that $T_1^s$ decreases monotonically with $\Delta\tau=\tau_2-\tau_1$, which is demonstrated in Fig.~\ref{fig:temp}. Note that with increasing $\kappa_1$ and $\kappa_2$ fixed at $0$, the decrease in $T_1^s$ becomes gradually slower, and at a high-enough value of $\kappa_1$, $T_1^s$ becomes almost constant with $\Delta \tau$.

\subsubsection{Effects of four-body interaction is negligible}
\label{subsubsection:four-body}
\begin{figure*}
\includegraphics[width=\textwidth]{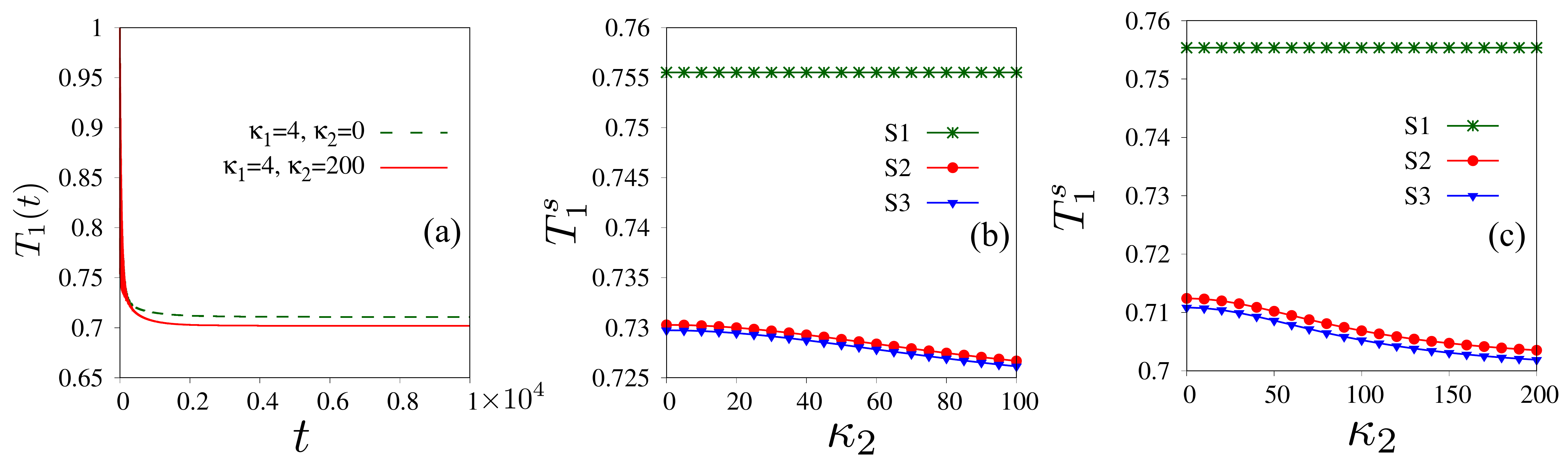}
\caption{\textbf{Response to four-body interaction term is insignificant.} We analyze here the consequences of a four-body interaction term in the Hamiltonian representing the interaction between the bath $B_1$ and the qubits $1$ and $2$. For $a)$, we depict the temperature of cold qubit, $T_1(t)$, vs the scaled time. The operating regime of the machine is chosen to be the strong-coupling regime ($g=0.5$), and the values of the qubit-bath interaction parameters are $\delta_1=10^{-4},\delta_2=10^{-8}$. For $b)$ and $c)$ the operating regime is weak-coupling $(g=10^{-2})$ and strong-coupling $(g=0.5)$, respectively, and the values of the qubit-bath interaction parameters are the same as in Fig. \ref{fig:harmonic}. Trends of $T_1^S$ with the increase of coupling of four-body interaction term, $\kappa_2$, are shown. All quantities plotted are dimensionless.}
\label{fig:four_body}
\end{figure*}

At this point, a natural question is whether the steady-state temperature of the cold qubit continues to decrease as one keeps on adding higher order interaction terms in the qubit-bath interactions corresponding to the bath $B_1$. To investigate this question, we add a four-body interaction term of the form $H_{I,4}^{(1)}$ to $H_I^{(1)}$ (see Eq.~(\ref{eq:total_hamiltonian})). As already mentioned in Sec.~\ref{sec:model}, there are different possibilities for the form of the four-body interaction term obeying the energy conservation principle. We choose a specific term  among the different possibilities in order to check whether there is a substantial difference in results due to the presence of the four-body interaction in $H_I^{(1)}$. The strength of the four-body interaction term relative to $H_S+H_B+(H_{I,2}^{(1)}+H_{I,2}^{(2)})$ is taken to be $\kappa_2$. We find that for a fixed value of $\kappa_1\geq 0$, even  a very high value of $\kappa_2$ changes $T_1^s$ only negligibly. This trend is prominent only in the \textbf{S2}-\textbf{S3} type dynamics (see Figs.~\ref{fig:four_body}(b)-(c)). This observation is also supported by the variations of $T_1^s$ with $\Delta E$, $g$, and $\Delta\tau$, with $\kappa_2\neq 0$ (see Figs.~\ref{fig:calibration}-\ref{fig:temp}). A small effect of $H_{I,4}^{(1)}$ in the local temperature of the qubit $1$ can be understood from the fact that although $\kappa_2$, similar to $\kappa_1$, contributes as $\kappa_2^2$ in the expression of $\gamma_{1,4}(E)$, unlike $\gamma_{1,3}(E)$, $\gamma_{1,4}(E)$ depends on the Ohmic spectral function as $\sim h_1(E)h_1(E^\prime)\sim J_1(E)$. Therefore, the values of the transition rates $\gamma_{1,4}(E)$, within the validity of the rotating-wave approximation, become very small, and so is the effect of the four-body interaction term. Since the dependence of the transition rate $\gamma_{1,4}(E)$ on $\kappa_2$ and $J_1(E)$  remains the same for  all possible forms of the four-body interaction terms, the effect due to other forms of four-body interaction terms is also expected to be negligible.

Fig.~\ref{fig:temp} also brings forth a rather curious result. For $\kappa_1>0$, we get $T_1^s<\tau_1$ even for  negligible or  vanishing $\Delta\tau$; i.e., local cooling of qubit 1 occurs even without a temperature difference between the two baths. This finding motivates the question as to whether a local cooling of qubit 1 is possible if one works with a single bath and few-body interaction terms, which we briefly discuss in the succeeding subsection.

\subsubsection{Three-qubit refrigerator with a single bath?}
\begin{figure*}
\includegraphics[width=0.7\textwidth]{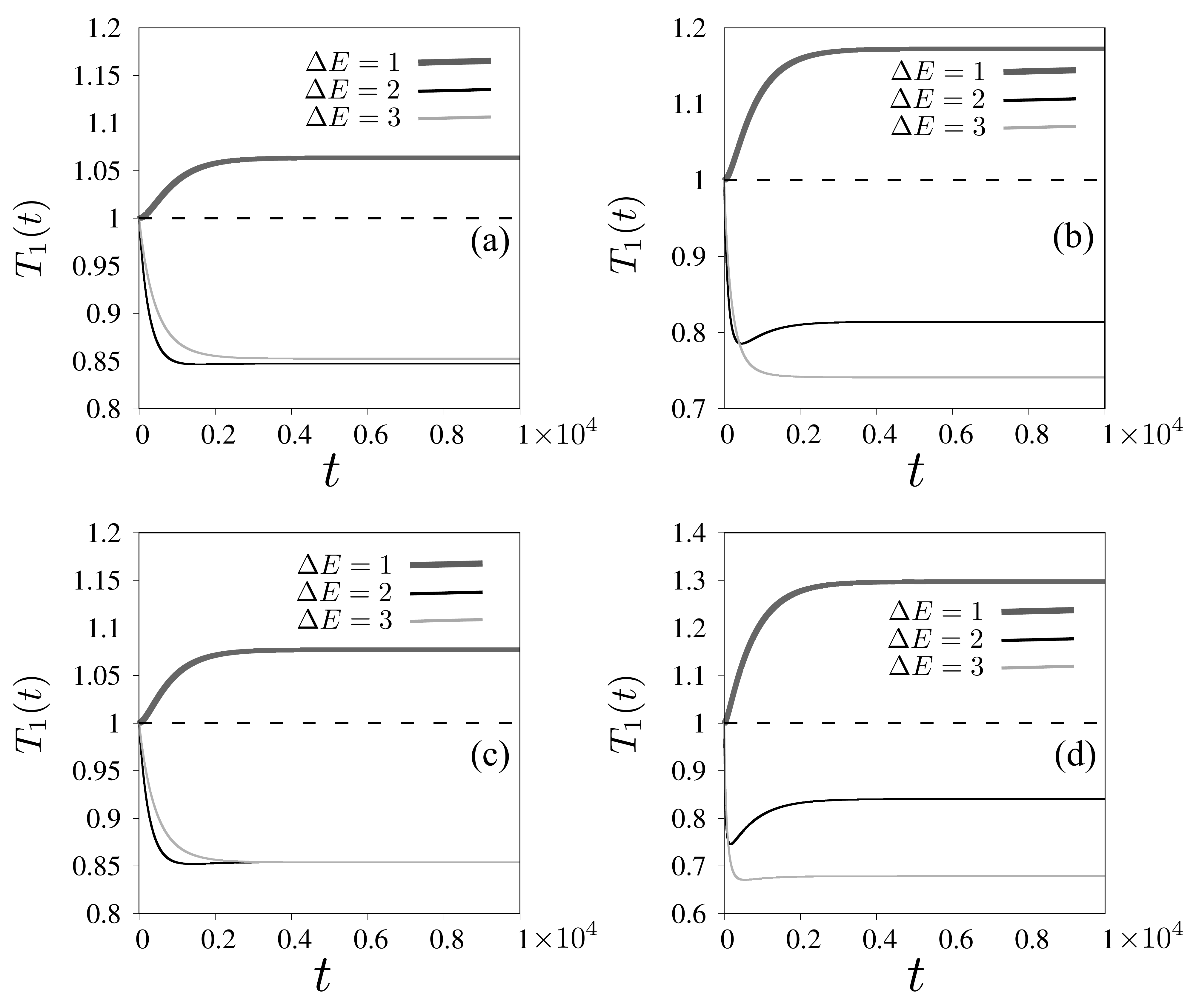}
\caption{\textbf{Local cooling in a three-qubit single-bath refrigerator.} We set the values of the parameters as $\delta=10^{-4}$ and $E_1=1$ for all panels. In each panel, we plot for three values of $\Delta E$, viz, \(1, 2, 3\).  
Weak-coupling instances ($g=10^{-2}$) are considered in panels $(a)$ for $\kappa_1=1,\kappa_2=0$ and $(b)$ for $\kappa_1=2,\kappa_2=0$, while strong-coupling ones ($g=0.5$) are in panels $(c)$ for $\kappa_1=1,\kappa_2=0$ and $(d)$ for $\kappa_1=4,\kappa_2=0$. The black dashed lines  represent the bath temperature $\tau=1$. All quantities plotted are dimensionless.
}
\label{fig:one_bath}
\end{figure*}

Here we assume that a system of three qubits is interacting with a common bosonic bath $B$ at temperature $\tau$, characterized by the Ohmic spectral function, $J(E)=\delta E\exp(-E/\Omega)$, where $\Omega$ is the cut-off frequency (see Sec.~\ref{sec:model}). The local Hamiltonian $H_0$ and the interaction Hamiltonian $H_1$ are as in Eqs. (\ref{eq:local-hamiltonian}) and (\ref{eq:int_Hamiltonian}), respectively, where we consider only two-body and three-body interaction terms for the qubit-bath interaction (i.e., $\kappa_2=0$). The Hamiltonian of the common bath $B$ is given by
\begin{eqnarray}
H_{B}=\int_{0}^{\omega}\tilde{\omega}\eta_{\omega^\prime}^{\dagger}\eta_{\omega^\prime}d\omega^\prime,
\end{eqnarray}
while the qubit-bath interaction Hamiltonian is 
\begin{eqnarray}
H_I=H_{I,2}+\kappa_1 H_{I,3},
\end{eqnarray}
with 
\begin{eqnarray}
H_{I,2}&=&\sum_{i=1}^3\int_{0}^{\omega}h(\omega)\sqrt{\tilde{\omega}}(\sigma_i^+ \eta_{\omega}+\sigma_i^- \eta_{\omega}^{\dagger})d\omega,\nonumber\\
H_{I,3}&=&\int_{0}^{\omega}h(\omega)\tilde{\omega}(\sigma_1^+\sigma_2^-+\sigma_1^-\sigma_2^+)\eta_{\omega}^{\dagger}\eta_{\omega}d\omega,\nonumber \\
&+&\int_{0}^{\omega}h(\omega)\tilde{\omega}(\sigma_2^+\sigma_3^-+\sigma_2^-\sigma_3^+)\eta_{\omega}^{\dagger}\eta_{\omega}d\omega,\nonumber \\
&+&\int_{0}^{\omega}h(\omega)\tilde{\omega}(\sigma_1^+\sigma_3^-+\sigma_1^-\sigma_3^+)\eta_{\omega}^{\dagger}\eta_{\omega}d\omega.
\end{eqnarray}
such that the total Hamiltonian for the system-environment duo is represented as
\begin{equation}
\tilde{H}=(H_0+gH_1)+H_{B}+H_{I,2}+\kappa_1H_{I,3}.
\end{equation}
In the presence of only two-body interaction between the qubits and the bath (i.e., with $\kappa_1=0$), none of the qubits is cooled, which is in contrast with the local cooling of  qubit 1 in the case of the three-qubit two-bath model (see Figs.~\ref{fig:harmonic}(a) and (d)). When $\kappa_1\neq 0$, local cooling of qubit 1 takes place, as depicted in Fig. \ref{fig:one_bath}. For both weak as well as strong coupling, an increase in the value of $\Delta E$ results in a local cooling of the first qubit. The thermodynamic consistency of this setup, as well as the setup with two baths of possibly equal temperatures,
is commented on in the following section.

\subsection{Thermodynamic consistency of single- and two-bath refrigerators}
\label{subsec:second_law}
For a quantum thermal machine described by a quantum master equation, the validity of the second law of thermodynamics is determined by the balance equation of entropy production rate,
\begin{equation}
\label{balance}
\frac{dS}{dt}-\sum_i \beta_iQ_i=\Delta
\end{equation}
where $S$ is the entropy of the system, $Q_i$ quantifies the flow of entropy (heat) from the system to the bath $i$, $\beta_i=\frac{1}{k_BT_i}$ with $k_B$ being the Boltzmann constant, $T_i$ is the absolute temperature of the bath $i$, and $\Delta$ is the source term quantifying the entropy production rate of the
system. It has been shown that the second law of thermodynamics is always valid for global master equations, but seems to be violated in cases of local master equations where the Lindblad operators are obtained by considering  transitions between  eigenstates of local Hamiltonians of the individual subsystems (e.g., $H_0$ in Eq.~(\ref{eq:local-hamiltonian}))~\cite{Barra2015,*Strasberg2017,*De_Chiara2018}. 
Therefore, the usual definition of the heat current, viz.  $Q_i=\text{Tr}(H_s \mathcal{L}_i(\rho_s))$, where $\mathcal{L}_i(\rho_s)$ is the dissipating term corresponding to the bath $i$ and $H_s$, is the local Hamiltonian, may not be a proper definition to check the validity of Eq.~(\ref{balance}), and the actual expression of the heat currents $Q_i$ has to be carefully determined~\cite{Hewgill2021}. 
In the single- and two-bath refrigerator setups, we have obtained the Lindblad operators by considering transitions between  eigenstates of the total system Hamiltonian, $H_S=H_0+H_1$, thereby following the global approach of the master equation. This implies that the balance equation is always valid in our case.
The excess terms that appear due to the consideration of the ``nonlocal'' term, \(H_1\), may also be  incorporated into the  entropy production rate term on the right-hand side of Eq.~(\ref{balance}), to be interpreted as ``excess source'' terms in the thermodynamics of the system. 
The weak-coupling Lindblad master equation sometimes fails to describe the stationary nonequilibrium properties, and this problem can also be resolved by constructing the master equation in a different way (see~\cite{Wichterich2007}).

\section{Refrigeration using reset model of qubit-bath interaction}
\label{sec:reset}
\begin{figure*}
\includegraphics[width=\textwidth]{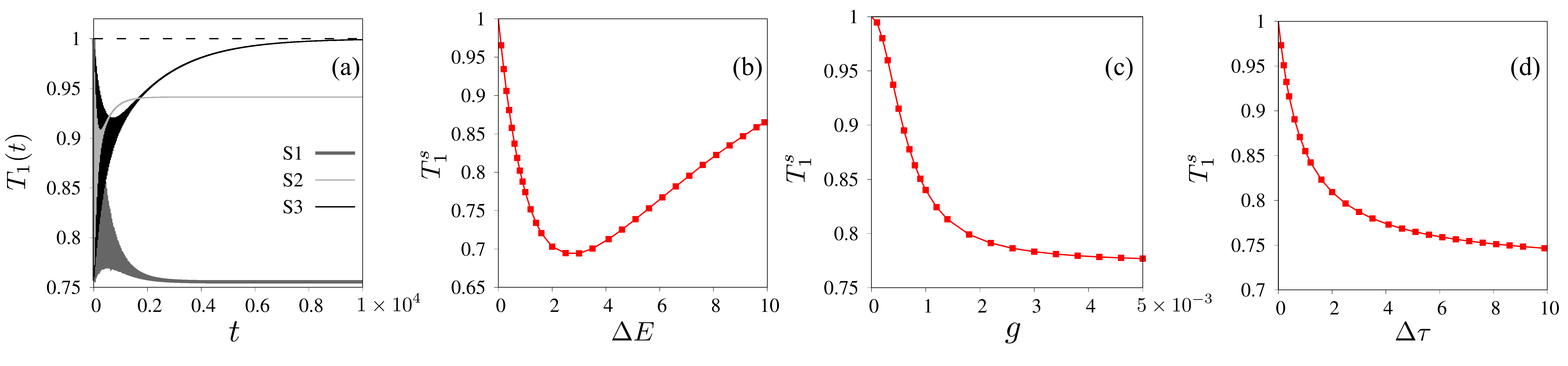}
\caption{\textbf{Consequences of using reset model.} (a) Dynamics of cold-qubit temperature in the weak-coupling ($g=10^{-2}$) operating regime. The values of the qubit-bath interaction parameters are chosen to be $i)\;\;\delta_1=10^{-7.5},\delta_2=10^{-3.5}$ for \textbf{S1}, $ii)\;\;\delta_1=10^{-3.5},\delta_2=10^{-7.5}$ for \textbf{S2}, and $iii)\;\;\delta_1=10^{-3},\delta_2=10^{-4}$ for \textbf{S3}. Variations of $T_1^s$ as  functions of (b) $\Delta E$ for $g=10^{-2}$, (c) $g$, and (d) $\Delta \tau$ for $g=10^{-2}$ are presented. The reset probabilities are taken to be $p_1=10^{-3.5}$ and $p_2=10^{-2.5}$ for (b), (c), and (d). The black dashed line in panel $(a)$ represent the cold bath temperature $\tau_1$. All quantities plotted are dimensionless.}
\label{fig:reset}
\end{figure*}

In order to check whether the features of the local steady-state cooling obtained in the three-qubit refrigerator model constituted using two thermal baths are artifacts of the specific qubit-bath interactions used in this paper, or are generic properties of the two-bath construction, we consider a second and rather well-studied model of qubit-bath interaction, namely, the reset model~\cite{linden2010,correa2013,mitchison2015,das2019}. We assume that at each time step, the bath induces a probabilistic reset on the attached qubit or group of qubits, such that the state of the qubit or the group of qubits is replaced by the initial ($t=0$) thermal state. In the two-bath  setup described in Fig.~\ref{fig:schematic}, the dynamical term $\mathcal{L}(\rho_S)$ (Eq.~(\ref{eq:qme})) can be written as (cf.~\cite{linden2010} for the three-bath setup)
\begin{eqnarray}
\mathcal{L}(\rho_S)&=&p_{1}\left[(\rho_1\otimes\rho_2)\otimes\text{Tr}_{12}(\rho_S(t))-\rho\right]\nonumber \\
&&+p_2\left[\rho_3\otimes\text{Tr}_{3}(\rho_S(t))-\rho\right],
\label{eq:reset}
\end{eqnarray}
where $\rho_1\otimes\rho_2$ and $\rho_{3}$ are respectively the thermal states of the qubit-pair $(1,2)$ and the qubit $3$ at $t=0$, and the states $\rho_i$ are given in Eq.~(\ref{eq:qubit-state}).  The qubit-bath interaction parameters $p_{\alpha}$, $\alpha=1,2$, are the probability densities per unit time, which provide the probability with which the reset operation is performed by the bath, and thereby provide a figure of merit of the quality of insulation of the qubit(s) from the bath(s). Here, we have assumed that the bath $B_1$ resets the states of the qubits $1$ and $2$ simultaneously. It can be shown by straightforward algebra that  the Lindblad operators corresponding to the dynamics described by Eq.~(\ref{eq:reset}) are of the form $L_{2,j}=\mathbb{I}_1\otimes\mathbb{I}_2\otimes A_{3,j}$  and  $L_{1,l}=A_{1,j}\otimes A_{2,k}\otimes\mathbb{I}_3$, with $j,k=1,2,3,4$ and $l=1,2,\cdots,16$, where the operators $\{A_{i,j}\}$ have the form $A_{i,1}=\sigma^-_i$, $A_{i,2}=\sigma^+_i$, $A_{i,3}=\frac{1}{2}(\mathbb{I}_i+\sigma_i^z)$, and $A_{i,4}=\frac{1}{2}(\mathbb{I}_i-\sigma_i^z)$. The transition rates $\gamma_{2,j}=p_2\tilde{r}_{2,j}$ (in Eq. (\ref{eq:lindblad_superoperator})) corresponding to the transformations due to the operators $L_{2,j}$ have two components, (i) the  qubit-bath interaction parameter $p_2$, and (ii) the temperature-dependent component $\{\tilde{r}_{2,j}\}$, $j=1,2,3,4$, with $\tilde{r}_{2,1}=1-r_3(0)$, $\tilde{r}_{2,2}=r_3(0)$, $\tilde{r}_{2,3}=r_3(0)$, and $\tilde{r}_{2,4}=1-r_3(0))$, which are fully determined by the thermal bath. In the case of the Lindblad operators $L_{1,l}$ corresponding to the qubit-pair $(1,2)$, the transition rates $\gamma_{1,l}$ can be obtained as $\gamma_{1,l}=p_{1}\tilde{r}_{a,j}\tilde{r}_{b,k}$, $l=1,2,\cdots,16$, $j,k=1,2,3,4$, where $\tilde{r}_{a,j}$ and $\tilde{r}_{b,k}$ have the same form as that of $\tilde{r}_{2,j}$. 

The discussion on the working principle of the three-qubit refrigerator (Sec.~\ref{subsec:working_principle}) is valid also for this model. However, it may not be intuitively clear whether the three types of dynamics of the temperature for the qubit $1$ are obtained once the qubit-qubit interaction is turned on. We answer this question affirmatively in this paper. In Fig~\ref{fig:reset}(a), we illustrate an example of the dynamics of the temperature of qubit $1$ in the three-qubit two-bath system (Fig.~\ref{fig:schematic}), where qubit $1$ has undergone dynamics of types \textbf{S1}, \textbf{S2},  and \textbf{S3} for $g=10^{-2}$. The calibration of the temperature of the steady state of qubit $1$ in the case of the reset model is qualitatively similar when $\Delta E$, $g$, and $\Delta\tau$ are chosen as the calibration parameter (see Figs.~\ref{fig:reset}(b)-(d)).  We point out here that the use of the reset model as the qubit-bath interaction has the liberty of using any value of the reset probability densities $p_i$, acting as the qubit-bath interaction parameters, which is in contrast to the case of the bosonic baths where the rotating-wave approximation imposes additional restrictions on the values of the system-, bath-, as well as the qubit-bath-interaction parameters. 
Therefore, a one-to-one comparison between the reset model and the model considering bosonic baths may not be possible.
\section{Discussion}
\label{sec:conclude}

We considered a quantum absorption refrigerator constituted of three interacting qubits and two thermal baths, where two of the qubits, including the target qubit for local cooling, are kept in contact with a common reservoir.  The interaction between each qubit and its corresponding bath is considered to be Markovian in nature, and we used thermal reservoirs made of an infinite number of quantum harmonic oscillators.  The use of the common reservoir results in the presence of three- and four-body interaction terms in the Hamiltonian representing the qubit-bath interactions. We specifically considered a three-body interaction term involving the interaction between the spin degree of freedom of the qubit with the number operator corresponding to a specific bosonic mode, and a four-body interaction term constituted of the spin degrees of freedom of the qubits and two different bath modes of the common bath. 

We first studied a situation where the few-body interactions are absent, and showed that the three-qubit two-bath  setup can indeed lead to a local steady-state cooling of one of the qubits connected to the common bath. We also showed that similar to a three-qubit three-bath refrigerator, a parameter regime exists also for the two-bath setup where little or no local steady-state cooling is achieved, thereby making the parameter regime disadvantageous for the functioning of the machine. However, in such situations, the few-body interaction terms play an important role. We demonstrated that in these parameter regions, upon turning on the strengths of the three- and four-body interaction terms, the steady-state temperature of the cold qubit decreases monotonically with an increase in the interaction strengths, resulting in substantial local steady-state cooling. Our findings suggest that the effect of the three-body interaction term in reducing the steady-state temperature is considerably higher compared to the four-body interaction term. We also explained the reason behind the effect of four- and higher than four-body interactions on the local steady-state cooling being so small compared to the three- and two-body terms. We further demonstrated that our setup can perform as a refrigerator even when the qubit-bath interaction is replaced by the widely studied reset model.

It is interesting to investigate the possible connections between the local temperature of the individual qubits and the different types of quantum correlations among the qubits in the system. We have checked  that in the three-qubit two-bath setup of the quantum refrigerator, bipartite entanglement~\cite{horodecki2009}, as measured by negativity~\cite{peres1996,horodecki1996,vidal2002}, in the partition $1:23$ of the three-qubit system remains at zero irrespective of the values of the system, the bath, and the system-bath interaction parameters as well as in the presence and absence of the few-body interaction terms in the qubit-bath interactions. The quantum coherence~\cite{Baumgratz2014} of qubit $1$, computed using the reduced state of qubit $1$ which is obtained by tracing out the rest of the qubits,  also exhibits similar behavior in the eigenbasis of the local Hamiltonian. However, the quantum coherence for  the three-qubit system in the computational multi-orthogonal product basis, 
and the mutual information~\cite{Witten2018} in the $1:23$ bipartition of the three-qubit system, oscillates at first, and then stabilizes to a nonzero value -- a behavior that is qualitatively similar to the dynamics of the total mutual information in the three-qubit three-bath setup~\cite{das2019}. These features also remain unchanged irrespective of whether the three- and four-body terms are present in the qubit-bath interactions.

\acknowledgements 

We acknowledge computations performed at the
cluster computing facility of Harish-Chandra Research Institute (HRI), India. S.D. acknowledges support from the Munich Center for Quantum Science and Technology. A.K.P. thanks HRI, India, for hospitality during his visits, and acknowledges IIT Palakkad for support from a seed Grant. The authors from HRI acknowledge support from the Department of Science and Technology, Government of India, through the QuEST grant (Grants No. DST/ICPS/QUST/Theme-1/2019/23 and No. DST/ICPS/QUST/Theme-3/2019/120). 

\appendix

\section{Explicit forms of Lindblad operators and transition rates}


\subsection{Lindblad operators}
\label{app:specific_lindblad}
In order to determine the explicit forms of the Lindblad operators, we note that the eigenstates of $H_S$ and the corresponding eigenenergies are given by Eqs. (\ref{equ:eigenstates}) and (\ref{equ:eigenenergies}). 
We first consider the Lindblad operators corresponding to the bath $B_1$ (i.e., for transitions corresponding to qubit-pair $12$). As per Eqs.~(\ref{eq:lindblad_general}) and (\ref{eq:all_states}), there are $18$ possible transition channels $L_{j,n}^E$ for the qubit pair $12$, among which the ones for positive energy gaps are given by 
\begin{eqnarray}
L_{1,2}^{E_1}&=&\ket{\mathbf{2}}\bra{\mathbf{0}}+\ket{\mathbf{5}}\bra{\mathbf{3}},\nonumber\\
L_{1,2}^{E_2}&=&\ket{\mathbf{3}}\bra{\mathbf{1}}+\ket{\mathbf{4}}\bra{\mathbf{2}},\nonumber\\
L_{1,2}^{\frac{1}{2}(E_{12}-E_3+\tilde{E})}&=&\epsilon_+\ket{\mathbf{4}}\bra{\mathbf{6}}+\sqrt{1-\epsilon_-^2}\ket{\mathbf{7}}\bra{\mathbf{1}},\nonumber\\
L_{1,2}^{\frac{1}{2}(E_{12}-E_3-\tilde{E})}&=&\epsilon_-\ket{\mathbf{4}}\bra{\mathbf{7}}+\sqrt{1-\epsilon_+^2}\ket{\mathbf{6}}\bra{\mathbf{1}},\nonumber\\
L_{1,2}^{\frac{1}{2}(E_{12}+E_3+\tilde{E})}&=&\epsilon_-\ket{\mathbf{7}}\bra{\mathbf{0}}+\sqrt{1+\epsilon_+^2}\ket{\mathbf{5}}\bra{\mathbf{6}},\nonumber\\
L_{1,2}^{\frac{1}{2}(E_{12}+E_3-\tilde{E})}&=&\epsilon_+\ket{\mathbf{6}}\bra{\mathbf{0}}+\sqrt{1-\epsilon_-^2}\ket{\mathbf{5}}\bra{\mathbf{7}},\nonumber\\
L_{1,3}^{\frac{1}{2}(E_3+\Delta E+\tilde{E})}&=&\epsilon_-\ket{\mathbf{7}}\bra{\mathbf{2}}+\sqrt{1-\epsilon_+^2}\ket{\mathbf{3}}\bra{\mathbf{6}},\nonumber\\
L_{1,3}^{\frac{1}{2}(E_3+\Delta E-\tilde{E})}&=&\epsilon_+\ket{\mathbf{6}}\bra{\mathbf{2}}+\sqrt{1-\epsilon_-^2}\ket{\mathbf{3}}\bra{\mathbf{7}},\nonumber\\
L_{1,4}^{E_{12}}&=&\ket{\mathbf{4}}\bra{\mathbf{0}}+\ket{\mathbf{5}}\bra{\mathbf{1}},
\end{eqnarray}
with $\epsilon_\pm=\varepsilon_\pm/\sqrt{1+|\varepsilon_\pm|^2}$.
The transition channels corresponding to the negative energy gaps, $-E$ , are obtained from $L_{1,n}^{-E}={L_{1,n}^{E}}^\dagger$, $n=2,3,4$. Similarly, there are $6$ possible transition channels for qubit $3$ connected to the bath $B_2$, and the ones corresponding to positive energy gaps are given by
\begin{eqnarray}
L_{2,2}^{E_3}&=&\ket{\mathbf{1}}\bra{\mathbf{0}}+\ket{\mathbf{5}}\bra{\mathbf{4}},\nonumber\\
L_{2,2}^{\frac{1}{2}(E_3+\Delta E+\tilde{E})}&=&\epsilon_+\ket{\mathbf{3}}\bra{\mathbf{6}}+\sqrt{1-\epsilon_-^2}\ket{\mathbf{7}}\bra{\mathbf{2}},\nonumber\\
L_{2,2}^{\frac{1}{2}(E_3+\Delta E+\tilde{E})}&=&\epsilon_-\ket{\mathbf{3}}\bra{\mathbf{7}}+\sqrt{1-\epsilon_+^2}\ket{\mathbf{6}}\bra{\mathbf{2}},
\end{eqnarray}
while the collapse operators corresponding to the negative energy gaps are given by $L_{2,2}^{-E}={L_{2,2}^{E}}^\dagger$.

\subsection{Transition rates}

The transition rates $\gamma(E)$ can be expressed in terms of the bath operators $b^{(j)}$ corresponding to the bath $B_j$ as~\cite{rivas2012}
\begin{equation}
\gamma_{j,n}^{kl}(E)=2\pi \mbox{Tr}\left[b^{(j)}_{k,n}(E)b^{(j)}_{l,n}\rho_{B_j}\right],
\label{gamma}
\end{equation}
with $\rho_{B_j}$ being the state of bath $B_j$ and $n=2,3,4$ corresponding to two-, three-, and four-body interaction terms.  Here, $k$ and $l$ are the indices that run from $1$ to $2$ and construct a $2\times 2$ matrix for each transition energy $E$. We denote the bath operators corresponding to the bath $B_j$ by $b_{l,n}^{(j)}$, while $b_{l,n}^{(j)}(E)$ is the bath operator for the transition energy $E$, with 
\begin{eqnarray}
b_{l,n}^{(j)}=\int_{-\omega_j}^{\omega_j} b_{l,n}^{(j)}(\omega) d\omega.
\label{eq:bath_op_definitions}
\end{eqnarray}
For the two-body interaction terms, $H_{I,2}^{(1)}$ and $H_{I,2}^{(2)}$, the bath operators, $b_{l,2}^{(j)}(E)$, are
\begin{eqnarray}
b_{1,2}^{(j)}(\omega)&=&
\begin{cases}
\frac{1}{2}\sqrt{\tilde{\omega}}h_i(\omega)\eta_{i,\omega} , & \omega > 0,\\
\frac{1}{2}\sqrt{\tilde{\omega}}h_i(\omega)\eta_{i,\omega}^{\dagger} , & \omega < 0,
\end{cases}\nonumber\\
b_{2,2}^{(j)}(\omega)&=&
\begin{cases}
-\frac{\text{i}}{2}\sqrt{\tilde{\omega}}h_i(\omega)\eta_{i,\omega} , & \omega > 0,\\
\frac{\text{i}}{2}\sqrt{\tilde{\omega}}h_i(\omega)\eta_{i,\omega}^{\dagger} , & \omega < 0,
\end{cases}
\end{eqnarray}
and the operators $b_{l,2}^{(j)}$ ($l=1,2$, $j=1,2$) are given by
\begin{eqnarray}
b_{l,2}^{(j)}&=&\frac{1}{2}\sqrt{\tilde{\omega}}\int_0^{\omega_j}h_j(\omega)(\eta_{i,\omega}+\eta_{j,\omega}^{\dagger})d\omega.
\end{eqnarray}
The corresponding transition rates, for $j=1,2$, are given by
\small 
\begin{eqnarray}
\gamma_{j,2}(E)&=&2\pi
\begin{cases}
J_{j}(E)[1+f(E,k_B^{-1}\tau_j^{-1})] , & E > 0,\\
J_{j}(|E|)f(|E|,k_B^{-1}\tau_j^{-1}), & E < 0.
\end{cases}
\label{eq:gamma2}
\end{eqnarray}\normalsize

Similarly, for the interaction term $H_{I,3}^{(1)}$, the bath operators are 
\begin{eqnarray}
b_{1,3}^{(1)}(\omega)&=&
\begin{cases}
\frac{1}{2}\kappa_1\tilde{\omega} h_1(\omega)\eta_{1,\omega}^{\dagger}\eta_{1,\omega} , & (\omega > 0)\\
\frac{1}{2}\kappa_1\tilde{\omega} h_1(\omega)\eta_{1,\omega}^{\dagger}\eta_{1,\omega}, & (\omega < 0)
\end{cases},\nonumber \\
b_{2,3}^{(1)}(\omega)&=&
\begin{cases}
-\frac{\text{i}}{2}\kappa_1\tilde{\omega} h_1(\omega)\eta_{1,\omega}^{\dagger}\eta_{1,\omega} , & (\omega > 0)\\
\frac{\text{i}}{2}\kappa_1\tilde{\omega} h_1(\omega)\eta_{1,\omega}^{\dagger}\eta_{1,\omega}, & (\omega < 0)
\end{cases}.
\end{eqnarray}
and 
\begin{eqnarray}
b_{1,3}^{(1)}&=&\kappa_1\tilde{\omega} \int_{0}^{\omega_1}d\omega  h_1(\omega)\eta_{1,\omega}^{\dagger}\eta_{1,\omega},\;
b_{2,3}^{(1)}
=0,
\end{eqnarray}
such that 
\begin{eqnarray}
\gamma_{1,3}(E)&=&2\pi \frac{\kappa_1^2}{\tilde{\omega}} \sqrt{J_1(|E|)} f(|E|,\beta_{1})\nonumber\\
&&\times\int_{0}^{\omega_1}d\omega \sqrt{J_1(\omega)} f(|\omega|,\beta_{1}).
\label{eq:gamma12}
\end{eqnarray}
So $\kappa_1^2$ can be represented as
\begin{eqnarray}
\kappa_1^2&=&\frac{\tilde{\omega}}{2\pi}\gamma_{1,3}(E) \frac{1}{\sqrt{J_1(|E|)}}f^{-1}(|E|,\beta_{1})\nonumber\\
&&\times\Big(\int_{0}^{\omega_1}d\omega \sqrt{J_1(\omega)} f(|\omega|,\beta_{1})\Big)^{-1}.
\label{eq:kappa1}
\end{eqnarray}
For the four-body interaction term $H_{I,4}^{(1)}$,  the bath operators $b_{l,4}^{(i)}(E)$ are
\small
\begin{eqnarray}
b_{1,4}^{(1)}(\omega)&=&
\begin{cases}
\frac{1}{2}\kappa_2\tilde{\omega} h_1(\omega) h_1(E_{12}-\omega)\eta_{1,\omega}\eta_{1,(E_{12}-\omega)} , & \omega > 0,\\
\frac{1}{2}\kappa_2\tilde{\omega} h_1(\omega)
h_1(E_{12}-\omega)\eta_{1,\omega}^{\dagger}\eta_{1,(E_{12}-\omega)}^{\dagger}, & \omega < 0,
\end{cases}\nonumber\\
b_{2,4}^{(1)}(\omega)&=&
\begin{cases}
-\frac{\text{i}}{2}\kappa_2\tilde{\omega} h_1(\omega)h_1(E_{12}-\omega)\eta_{1,\omega}\eta_{1,(E_{12}-\omega)} , & \omega > 0,\\
\frac{\text{i}}{2}\kappa_2\tilde{\omega} h_1(\omega)
h_1(E_{12}-\omega)\eta_{1,\omega}^{\dagger}\eta_{1,(E_{12}-\omega)}^{\dagger}, & \omega < 0,
\end{cases}\nonumber \\
\label{eq:bath_op_E}
\end{eqnarray}
\normalsize 
and the operators $b_{l,4}^{(j)}$ can be written as
\begin{widetext}
\begin{eqnarray}
b_{1,4}^{(1)}
&=&\frac{1}{2} \kappa_2\tilde{\omega} \int_{0}^{\omega_1}d\omega  h_1(\omega)h_1(E_{12}-\omega)\big[\eta_{1,\omega}\eta_{1,(E_{12}-\omega)}+\eta_{1,\omega}^{\dagger}\eta_{1,(E_{12}-\omega)}^{\dagger}\big],\nonumber \\
b_{2,4}^{(1)}&=&\frac{\text{i}}{2}\kappa_2\tilde{\omega} \int_{0}^{\omega_1}d\omega  h_1(\omega)h_1(E_{12}-\omega)\big[\eta_{1,\omega}\eta_{1,(E_{12}-\omega)}-\eta_{1,\omega}^{\dagger}\eta_{1,(E_{12}-\omega)}^{\dagger}\big],
\label{eq:bath_op_full}
\end{eqnarray}
\end{widetext}
where $l,j=1,2$. Substitution of Eqs.~(\ref{eq:bath_op_E}) and (\ref{eq:bath_op_full}) in Eq.~(\ref{gamma}) and subsequent algebra leads to 
\begin{widetext}
\begin{eqnarray}
\gamma_{1,4}(E)&=&2\pi \frac{\kappa_2^2}{\tilde{\omega}}
\begin{cases}
J_{1}(E)J_{1}(E_{12}-E)[1+f(E,k_B^{-1}\tau_1^{-1})][1+f(E_{12}-E,k_B^{-1}\tau_1^{-1})],& E > 0,\\
J_{1}(|E|)J_{1}(E_{12}-|E|)f(|E|,k_B^{-1}\tau_1^{-1})f(E_{12}-|E|,k_B^{-1}\tau_1^{-1}), & E < 0,
\end{cases} 
\label{eq:gamma11}
\end{eqnarray}
and $\kappa_2^2$ takes the form
\begin{eqnarray}
\kappa_2^2 &=& \frac{\tilde{\omega}}{2\pi}\gamma_{1,4}(E)
\begin{cases}
J_{1}^{-1}(E)J_{1}^{-1}(E_{12}-E)[1+f(E,k_B^{-1}\tau_1^{-1})]^{-1}[1+f(E_{12}-E,k_B^{-1}\tau_1^{-1})]^{-1},& E > 0,\\
J_{1}^{-1}(|E|)J_{1}^{-1}(E_{12}-|E|)f^{-1}(|E|,k_B^{-1}\tau_1^{-1})f^{-1}(E_{12}-|E|,k_B^{-1}\tau_1^{-1}), & E < 0,
\end{cases} 
\label{eq:kappa2}
\end{eqnarray}
\end{widetext}
where $f(E,k_B^{-1}\tau_{\alpha}^{-1})=1/(\exp(E/k_B\tau_\alpha)- 1)$ is the Bose-Einstein distribution. Note that in order to take care of the non-analyticity of $f(E_{12}-E,k_B^{-1}\tau_1^{-1})$ at $E=E_{12}$, we choose $h_1(E)$ such that $\lim_{E\rightarrow E_{12}}J(E_{12}-E)[1+f(E_{12}-E,k_B^{-1}\tau_1^{-1})]$ is finite. 

\bibliography{ref}{}

\begin{thebibliography}{88}%
\makeatletter
\providecommand \@ifxundefined [1]{%
 \@ifx{#1\undefined}
}%
\providecommand \@ifnum [1]{%
 \ifnum #1\expandafter \@firstoftwo
 \else \expandafter \@secondoftwo
 \fi
}%
\providecommand \@ifx [1]{%
 \ifx #1\expandafter \@firstoftwo
 \else \expandafter \@secondoftwo
 \fi
}%
\providecommand \natexlab [1]{#1}%
\providecommand \enquote  [1]{``#1''}%
\providecommand \bibnamefont  [1]{#1}%
\providecommand \bibfnamefont [1]{#1}%
\providecommand \citenamefont [1]{#1}%
\providecommand \href@noop [0]{\@secondoftwo}%
\providecommand \href [0]{\begingroup \@sanitize@url \@href}%
\providecommand \@href[1]{\@@startlink{#1}\@@href}%
\providecommand \@@href[1]{\endgroup#1\@@endlink}%
\providecommand \@sanitize@url [0]{\catcode `\\12\catcode `\$12\catcode
  `\&12\catcode `\#12\catcode `\^12\catcode `\_12\catcode `\%12\relax}%
\providecommand \@@startlink[1]{}%
\providecommand \@@endlink[0]{}%
\providecommand \url  [0]{\begingroup\@sanitize@url \@url }%
\providecommand \@url [1]{\endgroup\@href {#1}{\urlprefix }}%
\providecommand \urlprefix  [0]{URL }%
\providecommand \Eprint [0]{\href }%
\providecommand \doibase [0]{http://dx.doi.org/}%
\providecommand \selectlanguage [0]{\@gobble}%
\providecommand \bibinfo  [0]{\@secondoftwo}%
\providecommand \bibfield  [0]{\@secondoftwo}%
\providecommand \translation [1]{[#1]}%
\providecommand \BibitemOpen [0]{}%
\providecommand \bibitemStop [0]{}%
\providecommand \bibitemNoStop [0]{.\EOS\space}%
\providecommand \EOS [0]{\spacefactor3000\relax}%
\providecommand \BibitemShut  [1]{\csname bibitem#1\endcsname}%
\let\auto@bib@innerbib\@empty
\bibitem [{\citenamefont {Gemmer}\ \emph {et~al.}(2004)\citenamefont {Gemmer},
  \citenamefont {Michel},\ and\ \citenamefont {Mahler}}]{gemmer2004}%
  \BibitemOpen
  \bibfield  {author} {\bibinfo {author} {\bibfnamefont {G.}~\bibnamefont
  {Gemmer}}, \bibinfo {author} {\bibfnamefont {M.}~\bibnamefont {Michel}}, \
  and\ \bibinfo {author} {\bibfnamefont {G.}~\bibnamefont {Mahler}},\
  }\href@noop {} {\emph {\bibinfo {title} {Quantum Thermodynamics}}}\ (\bibinfo
   {publisher} {Springer, New York},\ \bibinfo {year} {2004})\BibitemShut
  {NoStop}%
\bibitem [{\citenamefont {Kosloff}(2013)}]{kosloff2013}%
  \BibitemOpen
  \bibfield  {author} {\bibinfo {author} {\bibfnamefont {R.}~\bibnamefont
  {Kosloff}},\ }\href {\doibase 10.3390/e15062100} {\bibfield  {journal}
  {\bibinfo  {journal} {Entropy}\ }\textbf {\bibinfo {volume} {15}},\ \bibinfo
  {pages} {2100} (\bibinfo {year} {2013})}\BibitemShut {NoStop}%
\bibitem [{\citenamefont {Gelbwaser-Klimovsky}\ \emph
  {et~al.}(2015)\citenamefont {Gelbwaser-Klimovsky}, \citenamefont {Niedenzu},\
  and\ \citenamefont {Kurizki}}]{kilmovski2015}%
  \BibitemOpen
  \bibfield  {author} {\bibinfo {author} {\bibfnamefont {D.}~\bibnamefont
  {Gelbwaser-Klimovsky}}, \bibinfo {author} {\bibfnamefont {W.}~\bibnamefont
  {Niedenzu}}, \ and\ \bibinfo {author} {\bibfnamefont {G.}~\bibnamefont
  {Kurizki}},\ }\href {\doibase 10.1016/bs.aamop.2015.07.002} {\bibfield
  {journal} {\bibinfo  {journal} {Adv. At. Mol. Opt. Phys.}\ }\textbf {\bibinfo
  {volume} {64}},\ \bibinfo {pages} {329} (\bibinfo {year} {2015})}\BibitemShut
  {NoStop}%
\bibitem [{\citenamefont {Misra}\ \emph {et~al.}(2015)\citenamefont {Misra},
  \citenamefont {Singh}, \citenamefont {Bera},\ and\ \citenamefont
  {Rajagopal}}]{Misra2015}%
  \BibitemOpen
  \bibfield  {author} {\bibinfo {author} {\bibfnamefont {A.}~\bibnamefont
  {Misra}}, \bibinfo {author} {\bibfnamefont {U.}~\bibnamefont {Singh}},
  \bibinfo {author} {\bibfnamefont {M.~N.}\ \bibnamefont {Bera}}, \ and\
  \bibinfo {author} {\bibfnamefont {A.~K.}\ \bibnamefont {Rajagopal}},\ }\href
  {\doibase 10.1103/PhysRevE.92.042161} {\bibfield  {journal} {\bibinfo
  {journal} {Phys. Rev. E}\ }\textbf {\bibinfo {volume} {92}},\ \bibinfo
  {pages} {042161} (\bibinfo {year} {2015})}\BibitemShut {NoStop}%
\bibitem [{\citenamefont {Millen}\ and\ \citenamefont
  {Xuereb}(2016)}]{millen2016}%
  \BibitemOpen
  \bibfield  {author} {\bibinfo {author} {\bibfnamefont {J.}~\bibnamefont
  {Millen}}\ and\ \bibinfo {author} {\bibfnamefont {A.}~\bibnamefont
  {Xuereb}},\ }\href {\doibase 10.1088/1367-2630/18/1/011002} {\bibfield
  {journal} {\bibinfo  {journal} {New Journal of Physics}\ }\textbf {\bibinfo
  {volume} {18}},\ \bibinfo {pages} {011002} (\bibinfo {year}
  {2016})}\BibitemShut {NoStop}%
\bibitem [{\citenamefont {Benenti}\ \emph {et~al.}(2017)\citenamefont
  {Benenti}, \citenamefont {Casati}, \citenamefont {Saito},\ and\ \citenamefont
  {Whitney}}]{benenti2017}%
  \BibitemOpen
  \bibfield  {author} {\bibinfo {author} {\bibfnamefont {G.}~\bibnamefont
  {Benenti}}, \bibinfo {author} {\bibfnamefont {G.}~\bibnamefont {Casati}},
  \bibinfo {author} {\bibfnamefont {K.}~\bibnamefont {Saito}}, \ and\ \bibinfo
  {author} {\bibfnamefont {R.~S.}\ \bibnamefont {Whitney}},\ }\href {\doibase
  10.1016/j.physrep.2017.05.008} {\bibfield  {journal} {\bibinfo  {journal}
  {Phys. Rep.}\ }\textbf {\bibinfo {volume} {694}},\ \bibinfo {pages} {1}
  (\bibinfo {year} {2017})}\BibitemShut {NoStop}%
\bibitem [{\citenamefont {Deffner}\ and\ \citenamefont
  {Campbell}(2019)}]{deffner2019}%
  \BibitemOpen
  \bibfield  {author} {\bibinfo {author} {\bibfnamefont {S.}~\bibnamefont
  {Deffner}}\ and\ \bibinfo {author} {\bibfnamefont {S.}~\bibnamefont
  {Campbell}},\ }\href {\doibase 10.1088/2053-2571/ab21c6} {\emph {\bibinfo
  {title} {Quantum Thermodynamics}}},\ 2053-2571\ (\bibinfo  {publisher}
  {Morgan and Claypool Publishers},\ \bibinfo {year} {2019})\BibitemShut
  {NoStop}%
\bibitem [{\citenamefont {Vinjanampathy}\ and\ \citenamefont
  {Anders}(2016)}]{vinjanampathy2016}%
  \BibitemOpen
  \bibfield  {author} {\bibinfo {author} {\bibfnamefont {S.}~\bibnamefont
  {Vinjanampathy}}\ and\ \bibinfo {author} {\bibfnamefont {J.}~\bibnamefont
  {Anders}},\ }\href {\doibase 10.1080/00107514.2016.1201896} {\bibfield
  {journal} {\bibinfo  {journal} {Contemporary Physics}\ }\textbf {\bibinfo
  {volume} {57}},\ \bibinfo {pages} {545} (\bibinfo {year} {2016})}\BibitemShut
  {NoStop}%
\bibitem [{\citenamefont {Goold}\ \emph {et~al.}(2016)\citenamefont {Goold},
  \citenamefont {Huber}, \citenamefont {Riera}, \citenamefont {del Rio},\ and\
  \citenamefont {Skrzypczyk}}]{goold2016}%
  \BibitemOpen
  \bibfield  {author} {\bibinfo {author} {\bibfnamefont {J.}~\bibnamefont
  {Goold}}, \bibinfo {author} {\bibfnamefont {M.}~\bibnamefont {Huber}},
  \bibinfo {author} {\bibfnamefont {A.}~\bibnamefont {Riera}}, \bibinfo
  {author} {\bibfnamefont {L.}~\bibnamefont {del Rio}}, \ and\ \bibinfo
  {author} {\bibfnamefont {P.}~\bibnamefont {Skrzypczyk}},\ }\href {\doibase
  10.1088/1751-8113/49/14/143001} {\bibfield  {journal} {\bibinfo  {journal}
  {Journal of Physics A: Mathematical and Theoretical}\ }\textbf {\bibinfo
  {volume} {49}},\ \bibinfo {pages} {143001} (\bibinfo {year}
  {2016})}\BibitemShut {NoStop}%
\bibitem [{\citenamefont {Allahverdyan}\ and\ \citenamefont
  {Nieuwenhuizen}(2000)}]{allahverdyan2000}%
  \BibitemOpen
  \bibfield  {author} {\bibinfo {author} {\bibfnamefont {A.~E.}\ \bibnamefont
  {Allahverdyan}}\ and\ \bibinfo {author} {\bibfnamefont {T.~M.}\ \bibnamefont
  {Nieuwenhuizen}},\ }\href {\doibase 10.1103/PhysRevLett.85.1799} {\bibfield
  {journal} {\bibinfo  {journal} {Phys. Rev. Lett.}\ }\textbf {\bibinfo
  {volume} {85}},\ \bibinfo {pages} {1799} (\bibinfo {year}
  {2000})}\BibitemShut {NoStop}%
\bibitem [{\citenamefont {Brand{\~a}o}\ \emph {et~al.}(2015)\citenamefont
  {Brand{\~a}o}, \citenamefont {Horodecki}, \citenamefont {Ng}, \citenamefont
  {Oppenheim},\ and\ \citenamefont {Wehner}}]{brandao2015}%
  \BibitemOpen
  \bibfield  {author} {\bibinfo {author} {\bibfnamefont {F.}~\bibnamefont
  {Brand{\~a}o}}, \bibinfo {author} {\bibfnamefont {M.}~\bibnamefont
  {Horodecki}}, \bibinfo {author} {\bibfnamefont {N.}~\bibnamefont {Ng}},
  \bibinfo {author} {\bibfnamefont {J.}~\bibnamefont {Oppenheim}}, \ and\
  \bibinfo {author} {\bibfnamefont {S.}~\bibnamefont {Wehner}},\ }\href
  {\doibase 10.1073/pnas.1411728112} {\bibfield  {journal} {\bibinfo  {journal}
  {Proceedings of the National Academy of Sciences}\ }\textbf {\bibinfo
  {volume} {112}},\ \bibinfo {pages} {3275} (\bibinfo {year}
  {2015})}\BibitemShut {NoStop}%
\bibitem [{\citenamefont {Gardas}\ and\ \citenamefont
  {Deffner}(2015)}]{gardas2015}%
  \BibitemOpen
  \bibfield  {author} {\bibinfo {author} {\bibfnamefont {B.}~\bibnamefont
  {Gardas}}\ and\ \bibinfo {author} {\bibfnamefont {S.}~\bibnamefont
  {Deffner}},\ }\href {\doibase 10.1103/PhysRevE.92.042126} {\bibfield
  {journal} {\bibinfo  {journal} {Phys. Rev. E}\ }\textbf {\bibinfo {volume}
  {92}},\ \bibinfo {pages} {042126} (\bibinfo {year} {2015})}\BibitemShut
  {NoStop}%
\bibitem [{\citenamefont {Palao}\ \emph {et~al.}(2001)\citenamefont {Palao},
  \citenamefont {Kosloff},\ and\ \citenamefont {Gordon}}]{palao2001}%
  \BibitemOpen
  \bibfield  {author} {\bibinfo {author} {\bibfnamefont {J.~P.}\ \bibnamefont
  {Palao}}, \bibinfo {author} {\bibfnamefont {R.}~\bibnamefont {Kosloff}}, \
  and\ \bibinfo {author} {\bibfnamefont {J.~M.}\ \bibnamefont {Gordon}},\
  }\href {\doibase 10.1103/PhysRevE.64.056130} {\bibfield  {journal} {\bibinfo
  {journal} {Phys. Rev. E}\ }\textbf {\bibinfo {volume} {64}},\ \bibinfo
  {pages} {056130} (\bibinfo {year} {2001})}\BibitemShut {NoStop}%
\bibitem [{\citenamefont {Feldmann}\ and\ \citenamefont
  {Kosloff}(2003)}]{feldmann2003}%
  \BibitemOpen
  \bibfield  {author} {\bibinfo {author} {\bibfnamefont {T.}~\bibnamefont
  {Feldmann}}\ and\ \bibinfo {author} {\bibfnamefont {R.}~\bibnamefont
  {Kosloff}},\ }\href {\doibase 10.1103/PhysRevE.68.016101} {\bibfield
  {journal} {\bibinfo  {journal} {Phys. Rev. E}\ }\textbf {\bibinfo {volume}
  {68}},\ \bibinfo {pages} {016101} (\bibinfo {year} {2003})}\BibitemShut
  {NoStop}%
\bibitem [{\citenamefont {Nimmrichter}\ \emph {et~al.}(2018)\citenamefont
  {Nimmrichter}, \citenamefont {Roulet},\ and\ \citenamefont
  {Scarani}}]{nimmrichter2018}%
  \BibitemOpen
  \bibfield  {author} {\bibinfo {author} {\bibfnamefont {S.}~\bibnamefont
  {Nimmrichter}}, \bibinfo {author} {\bibfnamefont {A.}~\bibnamefont {Roulet}},
  \ and\ \bibinfo {author} {\bibfnamefont {V.}~\bibnamefont {Scarani}},\
  }\enquote {\bibinfo {title} {Quantum rotor engines},}\ in\ \href {\doibase
  10.1007/978-3-319-99046-0_6} {\emph {\bibinfo {booktitle} {Thermodynamics in
  the Quantum Regime: Fundamental Aspects and New Directions}}},\ \bibinfo
  {editor} {edited by\ \bibinfo {editor} {\bibfnamefont {F.}~\bibnamefont
  {Binder}}, \bibinfo {editor} {\bibfnamefont {L.~A.}\ \bibnamefont {Correa}},
  \bibinfo {editor} {\bibfnamefont {C.}~\bibnamefont {Gogolin}}, \bibinfo
  {editor} {\bibfnamefont {J.}~\bibnamefont {Anders}}, \ and\ \bibinfo {editor}
  {\bibfnamefont {G.}~\bibnamefont {Adesso}}}\ (\bibinfo  {publisher} {Springer
  International Publishing},\ \bibinfo {address} {Cham},\ \bibinfo {year}
  {2018})\ pp.\ \bibinfo {pages} {227--245}\BibitemShut {NoStop}%
\bibitem [{\citenamefont {Kosloff}\ and\ \citenamefont
  {Levy}(2014)}]{kosloff2014}%
  \BibitemOpen
  \bibfield  {author} {\bibinfo {author} {\bibfnamefont {R.}~\bibnamefont
  {Kosloff}}\ and\ \bibinfo {author} {\bibfnamefont {A.}~\bibnamefont {Levy}},\
  }\href {\doibase 10.1146/annurev-physchem-040513-103724} {\bibfield
  {journal} {\bibinfo  {journal} {Annual Review of Physical Chemistry}\
  }\textbf {\bibinfo {volume} {65}},\ \bibinfo {pages} {365} (\bibinfo {year}
  {2014})}\BibitemShut {NoStop}%
\bibitem [{\citenamefont {Uzdin}\ \emph {et~al.}(2015)\citenamefont {Uzdin},
  \citenamefont {Levy},\ and\ \citenamefont {Kosloff}}]{uzdin2015}%
  \BibitemOpen
  \bibfield  {author} {\bibinfo {author} {\bibfnamefont {R.}~\bibnamefont
  {Uzdin}}, \bibinfo {author} {\bibfnamefont {A.}~\bibnamefont {Levy}}, \ and\
  \bibinfo {author} {\bibfnamefont {R.}~\bibnamefont {Kosloff}},\ }\href
  {\doibase 10.1103/PhysRevX.5.031044} {\bibfield  {journal} {\bibinfo
  {journal} {Phys. Rev. X}\ }\textbf {\bibinfo {volume} {5}},\ \bibinfo {pages}
  {031044} (\bibinfo {year} {2015})}\BibitemShut {NoStop}%
\bibitem [{\citenamefont {Levy}\ and\ \citenamefont
  {Kosloff}(2012)}]{levy2012}%
  \BibitemOpen
  \bibfield  {author} {\bibinfo {author} {\bibfnamefont {A.}~\bibnamefont
  {Levy}}\ and\ \bibinfo {author} {\bibfnamefont {R.}~\bibnamefont {Kosloff}},\
  }\href {\doibase 10.1103/PhysRevLett.108.070604} {\bibfield  {journal}
  {\bibinfo  {journal} {Phys. Rev. Lett.}\ }\textbf {\bibinfo {volume} {108}},\
  \bibinfo {pages} {070604} (\bibinfo {year} {2012})}\BibitemShut {NoStop}%
\bibitem [{\citenamefont {Clivaz}\ \emph {et~al.}(2019)\citenamefont {Clivaz},
  \citenamefont {Silva}, \citenamefont {Haack}, \citenamefont {Brask},
  \citenamefont {Brunner},\ and\ \citenamefont {Huber}}]{clivaz2019}%
  \BibitemOpen
  \bibfield  {author} {\bibinfo {author} {\bibfnamefont {F.}~\bibnamefont
  {Clivaz}}, \bibinfo {author} {\bibfnamefont {R.}~\bibnamefont {Silva}},
  \bibinfo {author} {\bibfnamefont {G.}~\bibnamefont {Haack}}, \bibinfo
  {author} {\bibfnamefont {J.~B.}\ \bibnamefont {Brask}}, \bibinfo {author}
  {\bibfnamefont {N.}~\bibnamefont {Brunner}}, \ and\ \bibinfo {author}
  {\bibfnamefont {M.}~\bibnamefont {Huber}},\ }\href {\doibase
  10.1103/PhysRevLett.123.170605} {\bibfield  {journal} {\bibinfo  {journal}
  {Phys. Rev. Lett.}\ }\textbf {\bibinfo {volume} {123}},\ \bibinfo {pages}
  {170605} (\bibinfo {year} {2019})}\BibitemShut {NoStop}%
\bibitem [{\citenamefont {Mitchison}(2019)}]{mitchison2019}%
  \BibitemOpen
  \bibfield  {author} {\bibinfo {author} {\bibfnamefont {M.~T.}\ \bibnamefont
  {Mitchison}},\ }\href {\doibase 10.1080/00107514.2019.1631555} {\bibfield
  {journal} {\bibinfo  {journal} {Contemporary Physics}\ }\textbf {\bibinfo
  {volume} {60}},\ \bibinfo {pages} {164} (\bibinfo {year} {2019})}\BibitemShut
  {NoStop}%
\bibitem [{\citenamefont {Bhattacharjee}\ and\ \citenamefont
  {Dutta}()}]{Bhattacharjee2020}%
  \BibitemOpen
  \bibfield  {author} {\bibinfo {author} {\bibfnamefont {S.}~\bibnamefont
  {Bhattacharjee}}\ and\ \bibinfo {author} {\bibfnamefont {A.}~\bibnamefont
  {Dutta}},\ }\href@noop {} {\bibinfo  {journal} {arXiv:2008.07889}\
  }\BibitemShut {NoStop}%
\bibitem [{\citenamefont {Geva}\ and\ \citenamefont
  {Kosloff}(1992)}]{geva1992}%
  \BibitemOpen
\bibfield  {journal} {  }\bibfield  {author} {\bibinfo {author} {\bibfnamefont
  {E.}~\bibnamefont {Geva}}\ and\ \bibinfo {author} {\bibfnamefont
  {R.}~\bibnamefont {Kosloff}},\ }\href {\doibase 10.1063/1.463909} {\bibfield
  {journal} {\bibinfo  {journal} {J. Chem. Phys.}\ }\textbf {\bibinfo {volume}
  {97}},\ \bibinfo {pages} {4398} (\bibinfo {year} {1992})}\BibitemShut
  {NoStop}%
\bibitem [{\citenamefont {Feldmann}\ and\ \citenamefont
  {Kosloff}(2000)}]{feldmann2000}%
  \BibitemOpen
  \bibfield  {author} {\bibinfo {author} {\bibfnamefont {T.}~\bibnamefont
  {Feldmann}}\ and\ \bibinfo {author} {\bibfnamefont {R.}~\bibnamefont
  {Kosloff}},\ }\href {\doibase 10.1103/PhysRevE.61.4774} {\bibfield  {journal}
  {\bibinfo  {journal} {Phys. Rev. E}\ }\textbf {\bibinfo {volume} {61}},\
  \bibinfo {pages} {4774} (\bibinfo {year} {2000})}\BibitemShut {NoStop}%
\bibitem [{\citenamefont {Ng}\ \emph {et~al.}(2017)\citenamefont {Ng},
  \citenamefont {Woods},\ and\ \citenamefont {Wehner}}]{ying2017}%
  \BibitemOpen
  \bibfield  {author} {\bibinfo {author} {\bibfnamefont {N.~H.~Y.}\
  \bibnamefont {Ng}}, \bibinfo {author} {\bibfnamefont {M.~P.}\ \bibnamefont
  {Woods}}, \ and\ \bibinfo {author} {\bibfnamefont {S.}~\bibnamefont
  {Wehner}},\ }\href {\doibase 10.1088/1367-2630/aa8ced} {\bibfield  {journal}
  {\bibinfo  {journal} {New Journal of Physics}\ }\textbf {\bibinfo {volume}
  {19}},\ \bibinfo {pages} {113005} (\bibinfo {year} {2017})}\BibitemShut
  {NoStop}%
\bibitem [{\citenamefont {Niedenzu}\ \emph {et~al.}(2018)\citenamefont
  {Niedenzu}, \citenamefont {Mukherjee}, \citenamefont {Ghosh}, \citenamefont
  {Kofman},\ and\ \citenamefont {Kurizki}}]{niedenzu2018}%
  \BibitemOpen
  \bibfield  {author} {\bibinfo {author} {\bibfnamefont {W.}~\bibnamefont
  {Niedenzu}}, \bibinfo {author} {\bibfnamefont {V.}~\bibnamefont {Mukherjee}},
  \bibinfo {author} {\bibfnamefont {A.}~\bibnamefont {Ghosh}}, \bibinfo
  {author} {\bibfnamefont {A.~G.}\ \bibnamefont {Kofman}}, \ and\ \bibinfo
  {author} {\bibfnamefont {G.}~\bibnamefont {Kurizki}},\ }\href {\doibase
  10.1038/s41467-017-01991-6} {\bibfield  {journal} {\bibinfo  {journal}
  {Nature Communications}\ }\textbf {\bibinfo {volume} {9}},\ \bibinfo {pages}
  {165} (\bibinfo {year} {2018})}\BibitemShut {NoStop}%
\bibitem [{\citenamefont {Xu}\ \emph {et~al.}(2018)\citenamefont {Xu},
  \citenamefont {Chen},\ and\ \citenamefont {Liu}}]{xu2018}%
  \BibitemOpen
  \bibfield  {author} {\bibinfo {author} {\bibfnamefont {Y.~Y.}\ \bibnamefont
  {Xu}}, \bibinfo {author} {\bibfnamefont {B.}~\bibnamefont {Chen}}, \ and\
  \bibinfo {author} {\bibfnamefont {J.}~\bibnamefont {Liu}},\ }\href {\doibase
  10.1103/PhysRevE.97.022130} {\bibfield  {journal} {\bibinfo  {journal} {Phys.
  Rev. E}\ }\textbf {\bibinfo {volume} {97}},\ \bibinfo {pages} {022130}
  (\bibinfo {year} {2018})}\BibitemShut {NoStop}%
\bibitem [{\citenamefont {Huber}\ \emph {et~al.}(2015)\citenamefont {Huber},
  \citenamefont {Perarnau-Llobet}, \citenamefont {Hovhannisyan}, \citenamefont
  {Skrzypczyk}, \citenamefont {Klöckl}, \citenamefont {Brunner},\ and\
  \citenamefont {Ac{\'{\i}}n}}]{huber2015}%
  \BibitemOpen
  \bibfield  {author} {\bibinfo {author} {\bibfnamefont {M.}~\bibnamefont
  {Huber}}, \bibinfo {author} {\bibfnamefont {M.}~\bibnamefont
  {Perarnau-Llobet}}, \bibinfo {author} {\bibfnamefont {K.~V.}\ \bibnamefont
  {Hovhannisyan}}, \bibinfo {author} {\bibfnamefont {P.}~\bibnamefont
  {Skrzypczyk}}, \bibinfo {author} {\bibfnamefont {C.}~\bibnamefont {Klöckl}},
  \bibinfo {author} {\bibfnamefont {N.}~\bibnamefont {Brunner}}, \ and\
  \bibinfo {author} {\bibfnamefont {A.}~\bibnamefont {Ac{\'{\i}}n}},\ }\href
  {\doibase 10.1088/1367-2630/17/6/065008} {\bibfield  {journal} {\bibinfo
  {journal} {New Journal of Physics}\ }\textbf {\bibinfo {volume} {17}},\
  \bibinfo {pages} {065008} (\bibinfo {year} {2015})}\BibitemShut {NoStop}%
\bibitem [{\citenamefont {Lostaglio}\ \emph {et~al.}(2015)\citenamefont
  {Lostaglio}, \citenamefont {Jennings},\ and\ \citenamefont
  {Rudolph}}]{lostaglio2015}%
  \BibitemOpen
  \bibfield  {author} {\bibinfo {author} {\bibfnamefont {M.}~\bibnamefont
  {Lostaglio}}, \bibinfo {author} {\bibfnamefont {D.}~\bibnamefont {Jennings}},
  \ and\ \bibinfo {author} {\bibfnamefont {T.}~\bibnamefont {Rudolph}},\ }\href
  {\doibase 10.1038/ncomms7383} {\bibfield  {journal} {\bibinfo  {journal}
  {Nature Communications}\ }\textbf {\bibinfo {volume} {6}},\ \bibinfo {pages}
  {6383} (\bibinfo {year} {2015})}\BibitemShut {NoStop}%
\bibitem [{\citenamefont {Ikonen}\ \emph {et~al.}(2017)\citenamefont {Ikonen},
  \citenamefont {Salmilehto},\ and\ \citenamefont
  {M{\"o}tt{\"o}nen}}]{ikonen2017}%
  \BibitemOpen
  \bibfield  {author} {\bibinfo {author} {\bibfnamefont {J.}~\bibnamefont
  {Ikonen}}, \bibinfo {author} {\bibfnamefont {J.}~\bibnamefont {Salmilehto}},
  \ and\ \bibinfo {author} {\bibfnamefont {M.}~\bibnamefont
  {M{\"o}tt{\"o}nen}},\ }\href {\doibase 10.1038/s41534-017-0015-5} {\bibfield
  {journal} {\bibinfo  {journal} {npj Quantum Information}\ }\textbf {\bibinfo
  {volume} {3}},\ \bibinfo {pages} {17} (\bibinfo {year} {2017})}\BibitemShut
  {NoStop}%
\bibitem [{\citenamefont {Campisi}\ \emph {et~al.}(2015)\citenamefont
  {Campisi}, \citenamefont {Pekola},\ and\ \citenamefont
  {Fazio}}]{campisi2015}%
  \BibitemOpen
  \bibfield  {author} {\bibinfo {author} {\bibfnamefont {M.}~\bibnamefont
  {Campisi}}, \bibinfo {author} {\bibfnamefont {J.}~\bibnamefont {Pekola}}, \
  and\ \bibinfo {author} {\bibfnamefont {R.}~\bibnamefont {Fazio}},\ }\href
  {\doibase 10.1088/1367-2630/17/3/035012} {\bibfield  {journal} {\bibinfo
  {journal} {New Journal of Physics}\ }\textbf {\bibinfo {volume} {17}},\
  \bibinfo {pages} {035012} (\bibinfo {year} {2015})}\BibitemShut {NoStop}%
\bibitem [{\citenamefont {D'Alessio}\ \emph {et~al.}(2016)\citenamefont
  {D'Alessio}, \citenamefont {Kafri}, \citenamefont {Polkovnikov},\ and\
  \citenamefont {Rigol}}]{dalessio2016}%
  \BibitemOpen
  \bibfield  {author} {\bibinfo {author} {\bibfnamefont {L.}~\bibnamefont
  {D'Alessio}}, \bibinfo {author} {\bibfnamefont {Y.}~\bibnamefont {Kafri}},
  \bibinfo {author} {\bibfnamefont {A.}~\bibnamefont {Polkovnikov}}, \ and\
  \bibinfo {author} {\bibfnamefont {M.}~\bibnamefont {Rigol}},\ }\href
  {\doibase 10.1080/00018732.2016.1198134} {\bibfield  {journal} {\bibinfo
  {journal} {Advances in Physics}\ }\textbf {\bibinfo {volume} {65}},\ \bibinfo
  {pages} {239} (\bibinfo {year} {2016})}\BibitemShut {NoStop}%
\bibitem [{\citenamefont {Gour}\ \emph {et~al.}(2015)\citenamefont {Gour},
  \citenamefont {M\"{u}ller}, \citenamefont {Narasimhachar}, \citenamefont
  {Spekkens},\ and\ \citenamefont {Halpern}}]{gour2015}%
  \BibitemOpen
  \bibfield  {author} {\bibinfo {author} {\bibfnamefont {G.}~\bibnamefont
  {Gour}}, \bibinfo {author} {\bibfnamefont {M.~P.}\ \bibnamefont
  {M\"{u}ller}}, \bibinfo {author} {\bibfnamefont {V.}~\bibnamefont
  {Narasimhachar}}, \bibinfo {author} {\bibfnamefont {R.~W.}\ \bibnamefont
  {Spekkens}}, \ and\ \bibinfo {author} {\bibfnamefont {N.~Y.}\ \bibnamefont
  {Halpern}},\ }\href {\doibase 10.1016/j.physrep.2015.04.003} {\bibfield
  {journal} {\bibinfo  {journal} {Phys. Rep.}\ }\textbf {\bibinfo {volume}
  {583}},\ \bibinfo {pages} {1} (\bibinfo {year} {2015})}\BibitemShut {NoStop}%
\bibitem [{\citenamefont {Dorner}\ \emph {et~al.}(2012)\citenamefont {Dorner},
  \citenamefont {Goold}, \citenamefont {Cormick}, \citenamefont {Paternostro},\
  and\ \citenamefont {Vedral}}]{dorner2012}%
  \BibitemOpen
  \bibfield  {author} {\bibinfo {author} {\bibfnamefont {R.}~\bibnamefont
  {Dorner}}, \bibinfo {author} {\bibfnamefont {J.}~\bibnamefont {Goold}},
  \bibinfo {author} {\bibfnamefont {C.}~\bibnamefont {Cormick}}, \bibinfo
  {author} {\bibfnamefont {M.}~\bibnamefont {Paternostro}}, \ and\ \bibinfo
  {author} {\bibfnamefont {V.}~\bibnamefont {Vedral}},\ }\href {\doibase
  10.1103/PhysRevLett.109.160601} {\bibfield  {journal} {\bibinfo  {journal}
  {Phys. Rev. Lett.}\ }\textbf {\bibinfo {volume} {109}},\ \bibinfo {pages}
  {160601} (\bibinfo {year} {2012})}\BibitemShut {NoStop}%
\bibitem [{\citenamefont {Mehboudi}\ \emph {et~al.}(2015)\citenamefont
  {Mehboudi}, \citenamefont {Moreno-Cardoner}, \citenamefont {Chiara},\ and\
  \citenamefont {Sanpera}}]{mehboudi2015}%
  \BibitemOpen
  \bibfield  {author} {\bibinfo {author} {\bibfnamefont {M.}~\bibnamefont
  {Mehboudi}}, \bibinfo {author} {\bibfnamefont {M.}~\bibnamefont
  {Moreno-Cardoner}}, \bibinfo {author} {\bibfnamefont {G.~D.}\ \bibnamefont
  {Chiara}}, \ and\ \bibinfo {author} {\bibfnamefont {A.}~\bibnamefont
  {Sanpera}},\ }\href {\doibase 10.1088/1367-2630/17/5/055020} {\bibfield
  {journal} {\bibinfo  {journal} {New Journal of Physics}\ }\textbf {\bibinfo
  {volume} {17}},\ \bibinfo {pages} {055020} (\bibinfo {year}
  {2015})}\BibitemShut {NoStop}%
\bibitem [{\citenamefont {Reimann}(2015)}]{reimann2015}%
  \BibitemOpen
  \bibfield  {author} {\bibinfo {author} {\bibfnamefont {P.}~\bibnamefont
  {Reimann}},\ }\href {\doibase 10.1088/1367-2630/17/5/055025} {\bibfield
  {journal} {\bibinfo  {journal} {New Journal of Physics}\ }\textbf {\bibinfo
  {volume} {17}},\ \bibinfo {pages} {055025} (\bibinfo {year}
  {2015})}\BibitemShut {NoStop}%
\bibitem [{\citenamefont {Eisert}\ \emph {et~al.}(2015)\citenamefont {Eisert},
  \citenamefont {Friesdorf},\ and\ \citenamefont {Gogolin}}]{eisert2015}%
  \BibitemOpen
  \bibfield  {author} {\bibinfo {author} {\bibfnamefont {J.}~\bibnamefont
  {Eisert}}, \bibinfo {author} {\bibfnamefont {M.}~\bibnamefont {Friesdorf}}, \
  and\ \bibinfo {author} {\bibfnamefont {C.}~\bibnamefont {Gogolin}},\ }\href
  {\doibase 10.1038/nphys3215} {\bibfield  {journal} {\bibinfo  {journal}
  {Nature Physics}\ }\textbf {\bibinfo {volume} {11}},\ \bibinfo {pages} {124}
  (\bibinfo {year} {2015})}\BibitemShut {NoStop}%
\bibitem [{\citenamefont {Gogolin}\ and\ \citenamefont
  {Eisert}(2016)}]{gogolin2016}%
  \BibitemOpen
  \bibfield  {author} {\bibinfo {author} {\bibfnamefont {C.}~\bibnamefont
  {Gogolin}}\ and\ \bibinfo {author} {\bibfnamefont {J.}~\bibnamefont
  {Eisert}},\ }\href {\doibase 10.1088/0034-4885/79/5/056001} {\bibfield
  {journal} {\bibinfo  {journal} {Reports on Progress in Physics}\ }\textbf
  {\bibinfo {volume} {79}},\ \bibinfo {pages} {056001} (\bibinfo {year}
  {2016})}\BibitemShut {NoStop}%
\bibitem [{\citenamefont {Skelt}\ \emph {et~al.}(2019)\citenamefont {Skelt},
  \citenamefont {Zawadzki},\ and\ \citenamefont {D'Amico}}]{skelt2019}%
  \BibitemOpen
  \bibfield  {author} {\bibinfo {author} {\bibfnamefont {A.~H.}\ \bibnamefont
  {Skelt}}, \bibinfo {author} {\bibfnamefont {K.}~\bibnamefont {Zawadzki}}, \
  and\ \bibinfo {author} {\bibfnamefont {I.}~\bibnamefont {D'Amico}},\ }\href
  {\doibase 10.1088/1751-8121/ab4fb6} {\bibfield  {journal} {\bibinfo
  {journal} {Journal of Physics A: Mathematical and Theoretical}\ }\textbf
  {\bibinfo {volume} {52}},\ \bibinfo {pages} {485304} (\bibinfo {year}
  {2019})}\BibitemShut {NoStop}%
\bibitem [{\citenamefont {Giazotto}\ \emph {et~al.}(2006)\citenamefont
  {Giazotto}, \citenamefont {Heikkil\"a}, \citenamefont {Luukanen},
  \citenamefont {Savin},\ and\ \citenamefont {Pekola}}]{giazotto2006}%
  \BibitemOpen
  \bibfield  {author} {\bibinfo {author} {\bibfnamefont {F.}~\bibnamefont
  {Giazotto}}, \bibinfo {author} {\bibfnamefont {T.~T.}\ \bibnamefont
  {Heikkil\"a}}, \bibinfo {author} {\bibfnamefont {A.}~\bibnamefont
  {Luukanen}}, \bibinfo {author} {\bibfnamefont {A.~M.}\ \bibnamefont {Savin}},
  \ and\ \bibinfo {author} {\bibfnamefont {J.~P.}\ \bibnamefont {Pekola}},\
  }\href {\doibase 10.1103/RevModPhys.78.217} {\bibfield  {journal} {\bibinfo
  {journal} {Rev. Mod. Phys.}\ }\textbf {\bibinfo {volume} {78}},\ \bibinfo
  {pages} {217} (\bibinfo {year} {2006})}\BibitemShut {NoStop}%
\bibitem [{\citenamefont {Abah}\ \emph {et~al.}(2012)\citenamefont {Abah},
  \citenamefont {Ro\ss{}nagel}, \citenamefont {Jacob}, \citenamefont {Deffner},
  \citenamefont {Schmidt-Kaler}, \citenamefont {Singer},\ and\ \citenamefont
  {Lutz}}]{abah2012}%
  \BibitemOpen
  \bibfield  {author} {\bibinfo {author} {\bibfnamefont {O.}~\bibnamefont
  {Abah}}, \bibinfo {author} {\bibfnamefont {J.}~\bibnamefont {Ro\ss{}nagel}},
  \bibinfo {author} {\bibfnamefont {G.}~\bibnamefont {Jacob}}, \bibinfo
  {author} {\bibfnamefont {S.}~\bibnamefont {Deffner}}, \bibinfo {author}
  {\bibfnamefont {F.}~\bibnamefont {Schmidt-Kaler}}, \bibinfo {author}
  {\bibfnamefont {K.}~\bibnamefont {Singer}}, \ and\ \bibinfo {author}
  {\bibfnamefont {E.}~\bibnamefont {Lutz}},\ }\href {\doibase
  10.1103/PhysRevLett.109.203006} {\bibfield  {journal} {\bibinfo  {journal}
  {Phys. Rev. Lett.}\ }\textbf {\bibinfo {volume} {109}},\ \bibinfo {pages}
  {203006} (\bibinfo {year} {2012})}\BibitemShut {NoStop}%
\bibitem [{\citenamefont {Ro{\ss}nagel}\ \emph {et~al.}(2016)\citenamefont
  {Ro{\ss}nagel}, \citenamefont {Dawkins}, \citenamefont {Tolazzi},
  \citenamefont {Abah}, \citenamefont {Lutz}, \citenamefont {Schmidt-Kaler},\
  and\ \citenamefont {Singer}}]{rossnage2016}%
  \BibitemOpen
  \bibfield  {author} {\bibinfo {author} {\bibfnamefont {J.}~\bibnamefont
  {Ro{\ss}nagel}}, \bibinfo {author} {\bibfnamefont {S.~T.}\ \bibnamefont
  {Dawkins}}, \bibinfo {author} {\bibfnamefont {K.~N.}\ \bibnamefont
  {Tolazzi}}, \bibinfo {author} {\bibfnamefont {O.}~\bibnamefont {Abah}},
  \bibinfo {author} {\bibfnamefont {E.}~\bibnamefont {Lutz}}, \bibinfo {author}
  {\bibfnamefont {F.}~\bibnamefont {Schmidt-Kaler}}, \ and\ \bibinfo {author}
  {\bibfnamefont {K.}~\bibnamefont {Singer}},\ }\href {\doibase
  10.1126/science.aad6320} {\bibfield  {journal} {\bibinfo  {journal}
  {Science}\ }\textbf {\bibinfo {volume} {352}},\ \bibinfo {pages} {325}
  (\bibinfo {year} {2016})}\BibitemShut {NoStop}%
\bibitem [{\citenamefont {Peterson}\ \emph {et~al.}(2019)\citenamefont
  {Peterson}, \citenamefont {Batalh\~ao}, \citenamefont {Herrera},
  \citenamefont {Souza}, \citenamefont {Sarthour}, \citenamefont {Oliveira},\
  and\ \citenamefont {Serra}}]{peterson2019}%
  \BibitemOpen
  \bibfield  {author} {\bibinfo {author} {\bibfnamefont {J.~P.~S.}\
  \bibnamefont {Peterson}}, \bibinfo {author} {\bibfnamefont {T.~B.}\
  \bibnamefont {Batalh\~ao}}, \bibinfo {author} {\bibfnamefont
  {M.}~\bibnamefont {Herrera}}, \bibinfo {author} {\bibfnamefont {A.~M.}\
  \bibnamefont {Souza}}, \bibinfo {author} {\bibfnamefont {R.~S.}\ \bibnamefont
  {Sarthour}}, \bibinfo {author} {\bibfnamefont {I.~S.}\ \bibnamefont
  {Oliveira}}, \ and\ \bibinfo {author} {\bibfnamefont {R.~M.}\ \bibnamefont
  {Serra}},\ }\href {\doibase 10.1103/PhysRevLett.123.240601} {\bibfield
  {journal} {\bibinfo  {journal} {Phys. Rev. Lett.}\ }\textbf {\bibinfo
  {volume} {123}},\ \bibinfo {pages} {240601} (\bibinfo {year}
  {2019})}\BibitemShut {NoStop}%
\bibitem [{\citenamefont {Karimi}\ and\ \citenamefont
  {Pekola}(2016)}]{karimi2016}%
  \BibitemOpen
  \bibfield  {author} {\bibinfo {author} {\bibfnamefont {B.}~\bibnamefont
  {Karimi}}\ and\ \bibinfo {author} {\bibfnamefont {J.~P.}\ \bibnamefont
  {Pekola}},\ }\href {\doibase 10.1103/PhysRevB.94.184503} {\bibfield
  {journal} {\bibinfo  {journal} {Phys. Rev. B}\ }\textbf {\bibinfo {volume}
  {94}},\ \bibinfo {pages} {184503} (\bibinfo {year} {2016})}\BibitemShut
  {NoStop}%
\bibitem [{\citenamefont {Hardal}\ \emph {et~al.}(2017)\citenamefont {Hardal},
  \citenamefont {Aslan}, \citenamefont {Wilson},\ and\ \citenamefont
  {M\"ustecapl\ifmmode \imath \else \i \fi{}o\ifmmode~\breve{g}\else
  \u{g}\fi{}lu}}]{hardal2017}%
  \BibitemOpen
  \bibfield  {author} {\bibinfo {author} {\bibfnamefont {A.~U.~C.}\
  \bibnamefont {Hardal}}, \bibinfo {author} {\bibfnamefont {N.}~\bibnamefont
  {Aslan}}, \bibinfo {author} {\bibfnamefont {C.~M.}\ \bibnamefont {Wilson}}, \
  and\ \bibinfo {author} {\bibfnamefont {O.~E.}\ \bibnamefont
  {M\"ustecapl\ifmmode \imath \else \i \fi{}o\ifmmode~\breve{g}\else
  \u{g}\fi{}lu}},\ }\href {\doibase 10.1103/PhysRevE.96.062120} {\bibfield
  {journal} {\bibinfo  {journal} {Phys. Rev. E}\ }\textbf {\bibinfo {volume}
  {96}},\ \bibinfo {pages} {062120} (\bibinfo {year} {2017})}\BibitemShut
  {NoStop}%
\bibitem [{\citenamefont {Manikandan}\ \emph {et~al.}(2019)\citenamefont
  {Manikandan}, \citenamefont {Giazotto},\ and\ \citenamefont
  {Jordan}}]{manikandan2019}%
  \BibitemOpen
  \bibfield  {author} {\bibinfo {author} {\bibfnamefont {S.~K.}\ \bibnamefont
  {Manikandan}}, \bibinfo {author} {\bibfnamefont {F.}~\bibnamefont
  {Giazotto}}, \ and\ \bibinfo {author} {\bibfnamefont {A.~N.}\ \bibnamefont
  {Jordan}},\ }\href {\doibase 10.1103/PhysRevApplied.11.054034} {\bibfield
  {journal} {\bibinfo  {journal} {Phys. Rev. Applied}\ }\textbf {\bibinfo
  {volume} {11}},\ \bibinfo {pages} {054034} (\bibinfo {year}
  {2019})}\BibitemShut {NoStop}%
\bibitem [{\citenamefont {Linden}\ \emph {et~al.}(2010)\citenamefont {Linden},
  \citenamefont {Popescu},\ and\ \citenamefont {Skrzypczyk}}]{linden2010}%
  \BibitemOpen
  \bibfield  {author} {\bibinfo {author} {\bibfnamefont {N.}~\bibnamefont
  {Linden}}, \bibinfo {author} {\bibfnamefont {S.}~\bibnamefont {Popescu}}, \
  and\ \bibinfo {author} {\bibfnamefont {P.}~\bibnamefont {Skrzypczyk}},\
  }\href {\doibase 10.1103/PhysRevLett.105.130401} {\bibfield  {journal}
  {\bibinfo  {journal} {Phys. Rev. Lett.}\ }\textbf {\bibinfo {volume} {105}},\
  \bibinfo {pages} {130401} (\bibinfo {year} {2010})}\BibitemShut {NoStop}%
\bibitem [{\citenamefont {Skrzypczyk}\ \emph {et~al.}(2011)\citenamefont
  {Skrzypczyk}, \citenamefont {Brunner}, \citenamefont {Linden},\ and\
  \citenamefont {Popescu}}]{skrzypczyk2011}%
  \BibitemOpen
  \bibfield  {author} {\bibinfo {author} {\bibfnamefont {P.}~\bibnamefont
  {Skrzypczyk}}, \bibinfo {author} {\bibfnamefont {N.}~\bibnamefont {Brunner}},
  \bibinfo {author} {\bibfnamefont {N.}~\bibnamefont {Linden}}, \ and\ \bibinfo
  {author} {\bibfnamefont {S.}~\bibnamefont {Popescu}},\ }\href {\doibase
  10.1088/1751-8113/44/49/492002} {\bibfield  {journal} {\bibinfo  {journal}
  {Journal of Physics A: Mathematical and Theoretical}\ }\textbf {\bibinfo
  {volume} {44}},\ \bibinfo {pages} {492002} (\bibinfo {year}
  {2011})}\BibitemShut {NoStop}%
\bibitem [{\citenamefont {Brunner}\ \emph {et~al.}(2012)\citenamefont
  {Brunner}, \citenamefont {Linden}, \citenamefont {Popescu},\ and\
  \citenamefont {Skrzypczyk}}]{brunner2012}%
  \BibitemOpen
  \bibfield  {author} {\bibinfo {author} {\bibfnamefont {N.}~\bibnamefont
  {Brunner}}, \bibinfo {author} {\bibfnamefont {N.}~\bibnamefont {Linden}},
  \bibinfo {author} {\bibfnamefont {S.}~\bibnamefont {Popescu}}, \ and\
  \bibinfo {author} {\bibfnamefont {P.}~\bibnamefont {Skrzypczyk}},\ }\href
  {\doibase 10.1103/PhysRevE.85.051117} {\bibfield  {journal} {\bibinfo
  {journal} {Phys. Rev. E}\ }\textbf {\bibinfo {volume} {85}},\ \bibinfo
  {pages} {051117} (\bibinfo {year} {2012})}\BibitemShut {NoStop}%
\bibitem [{\citenamefont {Brunner}\ \emph {et~al.}(2014)\citenamefont
  {Brunner}, \citenamefont {Huber}, \citenamefont {Linden}, \citenamefont
  {Popescu}, \citenamefont {Silva},\ and\ \citenamefont
  {Skrzypczyk}}]{brunner2014}%
  \BibitemOpen
  \bibfield  {author} {\bibinfo {author} {\bibfnamefont {N.}~\bibnamefont
  {Brunner}}, \bibinfo {author} {\bibfnamefont {M.}~\bibnamefont {Huber}},
  \bibinfo {author} {\bibfnamefont {N.}~\bibnamefont {Linden}}, \bibinfo
  {author} {\bibfnamefont {S.}~\bibnamefont {Popescu}}, \bibinfo {author}
  {\bibfnamefont {R.}~\bibnamefont {Silva}}, \ and\ \bibinfo {author}
  {\bibfnamefont {P.}~\bibnamefont {Skrzypczyk}},\ }\href {\doibase
  10.1103/PhysRevE.89.032115} {\bibfield  {journal} {\bibinfo  {journal} {Phys.
  Rev. E}\ }\textbf {\bibinfo {volume} {89}},\ \bibinfo {pages} {032115}
  (\bibinfo {year} {2014})}\BibitemShut {NoStop}%
\bibitem [{\citenamefont {Brask}\ and\ \citenamefont
  {Brunner}(2015)}]{brask2015}%
  \BibitemOpen
  \bibfield  {author} {\bibinfo {author} {\bibfnamefont {J.~B.}\ \bibnamefont
  {Brask}}\ and\ \bibinfo {author} {\bibfnamefont {N.}~\bibnamefont
  {Brunner}},\ }\href {\doibase 10.1103/PhysRevE.92.062101} {\bibfield
  {journal} {\bibinfo  {journal} {Phys. Rev. E}\ }\textbf {\bibinfo {volume}
  {92}},\ \bibinfo {pages} {062101} (\bibinfo {year} {2015})}\BibitemShut
  {NoStop}%
\bibitem [{\citenamefont {Correa}\ \emph {et~al.}(2013)\citenamefont {Correa},
  \citenamefont {Palao}, \citenamefont {Adesso},\ and\ \citenamefont
  {Alonso}}]{correa2013}%
  \BibitemOpen
  \bibfield  {author} {\bibinfo {author} {\bibfnamefont {L.~A.}\ \bibnamefont
  {Correa}}, \bibinfo {author} {\bibfnamefont {J.~P.}\ \bibnamefont {Palao}},
  \bibinfo {author} {\bibfnamefont {G.}~\bibnamefont {Adesso}}, \ and\ \bibinfo
  {author} {\bibfnamefont {D.}~\bibnamefont {Alonso}},\ }\href {\doibase
  10.1103/PhysRevE.87.042131} {\bibfield  {journal} {\bibinfo  {journal} {Phys.
  Rev. E}\ }\textbf {\bibinfo {volume} {87}},\ \bibinfo {pages} {042131}
  (\bibinfo {year} {2013})}\BibitemShut {NoStop}%
\bibitem [{\citenamefont {Correa}\ \emph {et~al.}(2014)\citenamefont {Correa},
  \citenamefont {Palao}, \citenamefont {Alonso},\ and\ \citenamefont
  {Adesso}}]{correa2014}%
  \BibitemOpen
  \bibfield  {author} {\bibinfo {author} {\bibfnamefont {L.~A.}\ \bibnamefont
  {Correa}}, \bibinfo {author} {\bibfnamefont {J.}~\bibnamefont {Palao}},
  \bibinfo {author} {\bibfnamefont {D.}~\bibnamefont {Alonso}}, \ and\ \bibinfo
  {author} {\bibfnamefont {G.}~\bibnamefont {Adesso}},\ }\href {\doibase
  10.1038/srep03949} {\bibfield  {journal} {\bibinfo  {journal} {Scientific
  Reports}\ }\textbf {\bibinfo {volume} {4}},\ \bibinfo {pages} {3949}
  (\bibinfo {year} {2014})}\BibitemShut {NoStop}%
\bibitem [{\citenamefont {Silva}\ \emph {et~al.}(2015)\citenamefont {Silva},
  \citenamefont {Skrzypczyk},\ and\ \citenamefont {Brunner}}]{silva2015}%
  \BibitemOpen
  \bibfield  {author} {\bibinfo {author} {\bibfnamefont {R.}~\bibnamefont
  {Silva}}, \bibinfo {author} {\bibfnamefont {P.}~\bibnamefont {Skrzypczyk}}, \
  and\ \bibinfo {author} {\bibfnamefont {N.}~\bibnamefont {Brunner}},\ }\href
  {\doibase 10.1103/PhysRevE.92.012136} {\bibfield  {journal} {\bibinfo
  {journal} {Phys. Rev. E}\ }\textbf {\bibinfo {volume} {92}},\ \bibinfo
  {pages} {012136} (\bibinfo {year} {2015})}\BibitemShut {NoStop}%
\bibitem [{\citenamefont {Erdman}\ \emph {et~al.}(2018)\citenamefont {Erdman},
  \citenamefont {Bhandari}, \citenamefont {Fazio}, \citenamefont {Pekola},\
  and\ \citenamefont {Taddei}}]{Erdman2018}%
  \BibitemOpen
  \bibfield  {author} {\bibinfo {author} {\bibfnamefont {P.~A.}\ \bibnamefont
  {Erdman}}, \bibinfo {author} {\bibfnamefont {B.}~\bibnamefont {Bhandari}},
  \bibinfo {author} {\bibfnamefont {R.}~\bibnamefont {Fazio}}, \bibinfo
  {author} {\bibfnamefont {J.~P.}\ \bibnamefont {Pekola}}, \ and\ \bibinfo
  {author} {\bibfnamefont {F.}~\bibnamefont {Taddei}},\ }\href {\doibase
  10.1103/PhysRevB.98.045433} {\bibfield  {journal} {\bibinfo  {journal} {Phys.
  Rev. B}\ }\textbf {\bibinfo {volume} {98}},\ \bibinfo {pages} {045433}
  (\bibinfo {year} {2018})}\BibitemShut {NoStop}%
\bibitem [{\citenamefont {Naseem}\ \emph {et~al.}(2020)\citenamefont {Naseem},
  \citenamefont {Misra},\ and\ \citenamefont {Özgür
  E~Müstecapl{\i}o{\u{g}}lu}}]{naseem2020}%
  \BibitemOpen
  \bibfield  {author} {\bibinfo {author} {\bibfnamefont {M.~T.}\ \bibnamefont
  {Naseem}}, \bibinfo {author} {\bibfnamefont {A.}~\bibnamefont {Misra}}, \
  and\ \bibinfo {author} {\bibnamefont {Özgür E~Müstecapl{\i}o{\u{g}}lu}},\
  }\href {\doibase 10.1088/2058-9565/ab8d89} {\bibfield  {journal} {\bibinfo
  {journal} {Quantum Science and Technology}\ }\textbf {\bibinfo {volume}
  {5}},\ \bibinfo {pages} {035006} (\bibinfo {year} {2020})}\BibitemShut
  {NoStop}%
\bibitem [{\citenamefont {Mitchison}\ \emph {et~al.}(2015)\citenamefont
  {Mitchison}, \citenamefont {Woods}, \citenamefont {Prior},\ and\
  \citenamefont {Huber}}]{mitchison2015}%
  \BibitemOpen
  \bibfield  {author} {\bibinfo {author} {\bibfnamefont {M.~T.}\ \bibnamefont
  {Mitchison}}, \bibinfo {author} {\bibfnamefont {M.~P.}\ \bibnamefont
  {Woods}}, \bibinfo {author} {\bibfnamefont {J.}~\bibnamefont {Prior}}, \ and\
  \bibinfo {author} {\bibfnamefont {M.}~\bibnamefont {Huber}},\ }\href
  {\doibase 10.1088/1367-2630/17/11/115013} {\bibfield  {journal} {\bibinfo
  {journal} {New Journal of Physics}\ }\textbf {\bibinfo {volume} {17}},\
  \bibinfo {pages} {115013} (\bibinfo {year} {2015})}\BibitemShut {NoStop}%
\bibitem [{\citenamefont {Das}\ \emph {et~al.}(2019)\citenamefont {Das},
  \citenamefont {Misra}, \citenamefont {Pal}, \citenamefont {Sen(De)},\ and\
  \citenamefont {Sen}}]{das2019}%
  \BibitemOpen
  \bibfield  {author} {\bibinfo {author} {\bibfnamefont {S.}~\bibnamefont
  {Das}}, \bibinfo {author} {\bibfnamefont {A.}~\bibnamefont {Misra}}, \bibinfo
  {author} {\bibfnamefont {A.~K.}\ \bibnamefont {Pal}}, \bibinfo {author}
  {\bibfnamefont {A.}~\bibnamefont {Sen(De)}}, \ and\ \bibinfo {author}
  {\bibfnamefont {U.}~\bibnamefont {Sen}},\ }\href {\doibase
  10.1209/0295-5075/125/20007} {\bibfield  {journal} {\bibinfo  {journal}
  {{EPL} (Europhysics Letters)}\ }\textbf {\bibinfo {volume} {125}},\ \bibinfo
  {pages} {20007} (\bibinfo {year} {2019})}\BibitemShut {NoStop}%
\bibitem [{\citenamefont {Man}\ and\ \citenamefont {Xia}(2017)}]{man2017}%
  \BibitemOpen
  \bibfield  {author} {\bibinfo {author} {\bibfnamefont {Z.-X.}\ \bibnamefont
  {Man}}\ and\ \bibinfo {author} {\bibfnamefont {Y.-J.}\ \bibnamefont {Xia}},\
  }\href {\doibase 10.1103/PhysRevE.96.012122} {\bibfield  {journal} {\bibinfo
  {journal} {Phys. Rev. E}\ }\textbf {\bibinfo {volume} {96}},\ \bibinfo
  {pages} {012122} (\bibinfo {year} {2017})}\BibitemShut {NoStop}%
\bibitem [{\citenamefont {Friedman}\ and\ \citenamefont
  {Segal}(2019)}]{friedman2019}%
  \BibitemOpen
  \bibfield  {author} {\bibinfo {author} {\bibfnamefont {H.~M.}\ \bibnamefont
  {Friedman}}\ and\ \bibinfo {author} {\bibfnamefont {D.}~\bibnamefont
  {Segal}},\ }\href {\doibase 10.1103/PhysRevE.100.062112} {\bibfield
  {journal} {\bibinfo  {journal} {Phys. Rev. E}\ }\textbf {\bibinfo {volume}
  {100}},\ \bibinfo {pages} {062112} (\bibinfo {year} {2019})}\BibitemShut
  {NoStop}%
\bibitem [{\citenamefont {Wang}\ \emph {et~al.}(2015)\citenamefont {Wang},
  \citenamefont {Lai}, \citenamefont {Ye}, \citenamefont {He}, \citenamefont
  {Ma},\ and\ \citenamefont {Liao}}]{wang2015}%
  \BibitemOpen
  \bibfield  {author} {\bibinfo {author} {\bibfnamefont {J.}~\bibnamefont
  {Wang}}, \bibinfo {author} {\bibfnamefont {Y.}~\bibnamefont {Lai}}, \bibinfo
  {author} {\bibfnamefont {Z.}~\bibnamefont {Ye}}, \bibinfo {author}
  {\bibfnamefont {J.}~\bibnamefont {He}}, \bibinfo {author} {\bibfnamefont
  {Y.}~\bibnamefont {Ma}}, \ and\ \bibinfo {author} {\bibfnamefont
  {Q.}~\bibnamefont {Liao}},\ }\href {\doibase 10.1103/PhysRevE.91.050102}
  {\bibfield  {journal} {\bibinfo  {journal} {Phys. Rev. E}\ }\textbf {\bibinfo
  {volume} {91}},\ \bibinfo {pages} {050102} (\bibinfo {year}
  {2015})}\BibitemShut {NoStop}%
\bibitem [{\citenamefont {He}\ \emph {et~al.}(2017)\citenamefont {He},
  \citenamefont {Huang},\ and\ \citenamefont {Yu}}]{he2017}%
  \BibitemOpen
  \bibfield  {author} {\bibinfo {author} {\bibfnamefont {Z.-c.}\ \bibnamefont
  {He}}, \bibinfo {author} {\bibfnamefont {X.-y.}\ \bibnamefont {Huang}}, \
  and\ \bibinfo {author} {\bibfnamefont {C.-s.}\ \bibnamefont {Yu}},\ }\href
  {\doibase 10.1103/PhysRevE.96.052126} {\bibfield  {journal} {\bibinfo
  {journal} {Phys. Rev. E}\ }\textbf {\bibinfo {volume} {96}},\ \bibinfo
  {pages} {052126} (\bibinfo {year} {2017})}\BibitemShut {NoStop}%
\bibitem [{\citenamefont {Du}\ and\ \citenamefont {Zhang}(2018)}]{du2018}%
  \BibitemOpen
  \bibfield  {author} {\bibinfo {author} {\bibfnamefont {J.-Y.}\ \bibnamefont
  {Du}}\ and\ \bibinfo {author} {\bibfnamefont {F.-L.}\ \bibnamefont {Zhang}},\
  }\href {\doibase 10.1088/1367-2630/aac688} {\bibfield  {journal} {\bibinfo
  {journal} {New Journal of Physics}\ }\textbf {\bibinfo {volume} {20}},\
  \bibinfo {pages} {063005} (\bibinfo {year} {2018})}\BibitemShut {NoStop}%
\bibitem [{\citenamefont {Mukhopadhyay}\ \emph {et~al.}(2018)\citenamefont
  {Mukhopadhyay}, \citenamefont {Misra}, \citenamefont {Bhattacharya},\ and\
  \citenamefont {Pati}}]{chiru2018}%
  \BibitemOpen
  \bibfield  {author} {\bibinfo {author} {\bibfnamefont {C.}~\bibnamefont
  {Mukhopadhyay}}, \bibinfo {author} {\bibfnamefont {A.}~\bibnamefont {Misra}},
  \bibinfo {author} {\bibfnamefont {S.}~\bibnamefont {Bhattacharya}}, \ and\
  \bibinfo {author} {\bibfnamefont {A.~K.}\ \bibnamefont {Pati}},\ }\href
  {\doibase 10.1103/PhysRevE.97.062116} {\bibfield  {journal} {\bibinfo
  {journal} {Phys. Rev. E}\ }\textbf {\bibinfo {volume} {97}},\ \bibinfo
  {pages} {062116} (\bibinfo {year} {2018})}\BibitemShut {NoStop}%
\bibitem [{\citenamefont {Seah}\ \emph {et~al.}(2018)\citenamefont {Seah},
  \citenamefont {Nimmrichter},\ and\ \citenamefont {Scarani}}]{seah2018}%
  \BibitemOpen
  \bibfield  {author} {\bibinfo {author} {\bibfnamefont {S.}~\bibnamefont
  {Seah}}, \bibinfo {author} {\bibfnamefont {S.}~\bibnamefont {Nimmrichter}}, \
  and\ \bibinfo {author} {\bibfnamefont {V.}~\bibnamefont {Scarani}},\ }\href
  {\doibase 10.1103/PhysRevE.98.012131} {\bibfield  {journal} {\bibinfo
  {journal} {Phys. Rev. E}\ }\textbf {\bibinfo {volume} {98}},\ \bibinfo
  {pages} {012131} (\bibinfo {year} {2018})}\BibitemShut {NoStop}%
\bibitem [{\citenamefont {Barra}\ and\ \citenamefont
  {Lled{\'o}}(2018)}]{barra2018}%
  \BibitemOpen
  \bibfield  {author} {\bibinfo {author} {\bibfnamefont {F.}~\bibnamefont
  {Barra}}\ and\ \bibinfo {author} {\bibfnamefont {C.}~\bibnamefont
  {Lled{\'o}}},\ }\href {\doibase 10.1140/epjst/e2018-00084-x} {\bibfield
  {journal} {\bibinfo  {journal} {The European Physical Journal Special
  Topics}\ }\textbf {\bibinfo {volume} {227}},\ \bibinfo {pages} {231}
  (\bibinfo {year} {2018})}\BibitemShut {NoStop}%
\bibitem [{\citenamefont {Hewgill}\ \emph {et~al.}(2020)\citenamefont
  {Hewgill}, \citenamefont {Gonz\'alez}, \citenamefont {Palao}, \citenamefont
  {Alonso}, \citenamefont {Ferraro},\ and\ \citenamefont
  {De~Chiara}}]{hewgill2020}%
  \BibitemOpen
  \bibfield  {author} {\bibinfo {author} {\bibfnamefont {A.}~\bibnamefont
  {Hewgill}}, \bibinfo {author} {\bibfnamefont {J.~O.}\ \bibnamefont
  {Gonz\'alez}}, \bibinfo {author} {\bibfnamefont {J.~P.}\ \bibnamefont
  {Palao}}, \bibinfo {author} {\bibfnamefont {D.}~\bibnamefont {Alonso}},
  \bibinfo {author} {\bibfnamefont {A.}~\bibnamefont {Ferraro}}, \ and\
  \bibinfo {author} {\bibfnamefont {G.}~\bibnamefont {De~Chiara}},\ }\href
  {\doibase 10.1103/PhysRevE.101.012109} {\bibfield  {journal} {\bibinfo
  {journal} {Phys. Rev. E}\ }\textbf {\bibinfo {volume} {101}},\ \bibinfo
  {pages} {012109} (\bibinfo {year} {2020})}\BibitemShut {NoStop}%
\bibitem [{\citenamefont {Venturelli}\ \emph {et~al.}(2013)\citenamefont
  {Venturelli}, \citenamefont {Fazio},\ and\ \citenamefont
  {Giovannetti}}]{venturelli2013}%
  \BibitemOpen
  \bibfield  {author} {\bibinfo {author} {\bibfnamefont {D.}~\bibnamefont
  {Venturelli}}, \bibinfo {author} {\bibfnamefont {R.}~\bibnamefont {Fazio}}, \
  and\ \bibinfo {author} {\bibfnamefont {V.}~\bibnamefont {Giovannetti}},\
  }\href {\doibase 10.1103/PhysRevLett.110.256801} {\bibfield  {journal}
  {\bibinfo  {journal} {Phys. Rev. Lett.}\ }\textbf {\bibinfo {volume} {110}},\
  \bibinfo {pages} {256801} (\bibinfo {year} {2013})}\BibitemShut {NoStop}%
\bibitem [{\citenamefont {Hofer}\ \emph {et~al.}(2016)\citenamefont {Hofer},
  \citenamefont {Perarnau-Llobet}, \citenamefont {Brask}, \citenamefont
  {Silva}, \citenamefont {Huber},\ and\ \citenamefont {Brunner}}]{hofer2016}%
  \BibitemOpen
  \bibfield  {author} {\bibinfo {author} {\bibfnamefont {P.~P.}\ \bibnamefont
  {Hofer}}, \bibinfo {author} {\bibfnamefont {M.}~\bibnamefont
  {Perarnau-Llobet}}, \bibinfo {author} {\bibfnamefont {J.~B.}\ \bibnamefont
  {Brask}}, \bibinfo {author} {\bibfnamefont {R.}~\bibnamefont {Silva}},
  \bibinfo {author} {\bibfnamefont {M.}~\bibnamefont {Huber}}, \ and\ \bibinfo
  {author} {\bibfnamefont {N.}~\bibnamefont {Brunner}},\ }\href {\doibase
  10.1103/PhysRevB.94.235420} {\bibfield  {journal} {\bibinfo  {journal} {Phys.
  Rev. B}\ }\textbf {\bibinfo {volume} {94}},\ \bibinfo {pages} {235420}
  (\bibinfo {year} {2016})}\BibitemShut {NoStop}%
\bibitem [{\citenamefont {Mitchison}\ \emph {et~al.}(2016)\citenamefont
  {Mitchison}, \citenamefont {Huber}, \citenamefont {Prior}, \citenamefont
  {Woods},\ and\ \citenamefont {Plenio}}]{mitchison2016}%
  \BibitemOpen
  \bibfield  {author} {\bibinfo {author} {\bibfnamefont {M.~T.}\ \bibnamefont
  {Mitchison}}, \bibinfo {author} {\bibfnamefont {M.}~\bibnamefont {Huber}},
  \bibinfo {author} {\bibfnamefont {J.}~\bibnamefont {Prior}}, \bibinfo
  {author} {\bibfnamefont {M.~P.}\ \bibnamefont {Woods}}, \ and\ \bibinfo
  {author} {\bibfnamefont {M.~B.}\ \bibnamefont {Plenio}},\ }\href {\doibase
  10.1088/2058-9565/1/1/015001} {\bibfield  {journal} {\bibinfo  {journal}
  {Quantum Science and Technology}\ }\textbf {\bibinfo {volume} {1}},\ \bibinfo
  {pages} {015001} (\bibinfo {year} {2016})}\BibitemShut {NoStop}%
\bibitem [{\citenamefont {Mitchison}\ and\ \citenamefont
  {Potts}(2018)}]{mitchison2018}%
  \BibitemOpen
  \bibfield  {author} {\bibinfo {author} {\bibfnamefont {M.~T.}\ \bibnamefont
  {Mitchison}}\ and\ \bibinfo {author} {\bibfnamefont {P.~P.}\ \bibnamefont
  {Potts}},\ }\enquote {\bibinfo {title} {Physical implementations of quantum
  absorption refrigerators},}\ in\ \href {\doibase 10.1007/978-3-319-99046-0_6}
  {\emph {\bibinfo {booktitle} {Thermodynamics in the Quantum Regime:
  Fundamental Aspects and New Directions}}},\ \bibinfo {editor} {edited by\
  \bibinfo {editor} {\bibfnamefont {F.}~\bibnamefont {Binder}}, \bibinfo
  {editor} {\bibfnamefont {L.~A.}\ \bibnamefont {Correa}}, \bibinfo {editor}
  {\bibfnamefont {C.}~\bibnamefont {Gogolin}}, \bibinfo {editor} {\bibfnamefont
  {J.}~\bibnamefont {Anders}}, \ and\ \bibinfo {editor} {\bibfnamefont
  {G.}~\bibnamefont {Adesso}}}\ (\bibinfo  {publisher} {Springer International
  Publishing},\ \bibinfo {address} {Cham},\ \bibinfo {year} {2018})\ pp.\
  \bibinfo {pages} {149--174}\BibitemShut {NoStop}%
\bibitem [{\citenamefont {Maslennikov}\ \emph {et~al.}(2019)\citenamefont
  {Maslennikov}, \citenamefont {Ding}, \citenamefont {Habl{\"u}tzel},
  \citenamefont {Gan}, \citenamefont {Roulet}, \citenamefont {Nimmrichter},
  \citenamefont {Dai}, \citenamefont {Scarani},\ and\ \citenamefont
  {Matsukevich}}]{maslennikov2019}%
  \BibitemOpen
  \bibfield  {author} {\bibinfo {author} {\bibfnamefont {G.}~\bibnamefont
  {Maslennikov}}, \bibinfo {author} {\bibfnamefont {S.}~\bibnamefont {Ding}},
  \bibinfo {author} {\bibfnamefont {R.}~\bibnamefont {Habl{\"u}tzel}}, \bibinfo
  {author} {\bibfnamefont {J.}~\bibnamefont {Gan}}, \bibinfo {author}
  {\bibfnamefont {A.}~\bibnamefont {Roulet}}, \bibinfo {author} {\bibfnamefont
  {S.}~\bibnamefont {Nimmrichter}}, \bibinfo {author} {\bibfnamefont
  {J.}~\bibnamefont {Dai}}, \bibinfo {author} {\bibfnamefont {V.}~\bibnamefont
  {Scarani}}, \ and\ \bibinfo {author} {\bibfnamefont {D.}~\bibnamefont
  {Matsukevich}},\ }\href {\doibase 10.1038/s41467-018-08090-0} {\bibfield
  {journal} {\bibinfo  {journal} {Nature Communications}\ }\textbf {\bibinfo
  {volume} {10}},\ \bibinfo {pages} {202} (\bibinfo {year} {2019})}\BibitemShut
  {NoStop}%
\bibitem [{\citenamefont {Basilewitsch}\ \emph {et~al.}(2019)\citenamefont
  {Basilewitsch}, \citenamefont {Cosco}, \citenamefont {Gullo}, \citenamefont
  {Möttönen}, \citenamefont {Ala-Nissilä}, \citenamefont {Koch},\ and\
  \citenamefont {Maniscalco}}]{basilewitsch2019}%
  \BibitemOpen
  \bibfield  {author} {\bibinfo {author} {\bibfnamefont {D.}~\bibnamefont
  {Basilewitsch}}, \bibinfo {author} {\bibfnamefont {F.}~\bibnamefont {Cosco}},
  \bibinfo {author} {\bibfnamefont {N.~L.}\ \bibnamefont {Gullo}}, \bibinfo
  {author} {\bibfnamefont {M.}~\bibnamefont {Möttönen}}, \bibinfo {author}
  {\bibfnamefont {T.}~\bibnamefont {Ala-Nissilä}}, \bibinfo {author}
  {\bibfnamefont {C.~P.}\ \bibnamefont {Koch}}, \ and\ \bibinfo {author}
  {\bibfnamefont {S.}~\bibnamefont {Maniscalco}},\ }\href {\doibase
  10.1088/1367-2630/ab41ad} {\bibfield  {journal} {\bibinfo  {journal} {New
  Journal of Physics}\ }\textbf {\bibinfo {volume} {21}},\ \bibinfo {pages}
  {093054} (\bibinfo {year} {2019})}\BibitemShut {NoStop}%
\bibitem [{\citenamefont {Poyatos}\ \emph {et~al.}(1996)\citenamefont
  {Poyatos}, \citenamefont {Cirac},\ and\ \citenamefont
  {Zoller}}]{poyatos1996}%
  \BibitemOpen
  \bibfield  {author} {\bibinfo {author} {\bibfnamefont {J.~F.}\ \bibnamefont
  {Poyatos}}, \bibinfo {author} {\bibfnamefont {J.~I.}\ \bibnamefont {Cirac}},
  \ and\ \bibinfo {author} {\bibfnamefont {P.}~\bibnamefont {Zoller}},\ }\href
  {\doibase 10.1103/PhysRevLett.77.4728} {\bibfield  {journal} {\bibinfo
  {journal} {Phys. Rev. Lett.}\ }\textbf {\bibinfo {volume} {77}},\ \bibinfo
  {pages} {4728} (\bibinfo {year} {1996})}\BibitemShut {NoStop}%
\bibitem [{\citenamefont {Myatt}\ \emph {et~al.}(2000)\citenamefont {Myatt},
  \citenamefont {King}, \citenamefont {Turchette}, \citenamefont {Sackett},
  \citenamefont {Kielpinski}, \citenamefont {Itano}, \citenamefont {Monroe},\
  and\ \citenamefont {Wineland}}]{myatt2000}%
  \BibitemOpen
  \bibfield  {author} {\bibinfo {author} {\bibfnamefont {C.~J.}\ \bibnamefont
  {Myatt}}, \bibinfo {author} {\bibfnamefont {B.~E.}\ \bibnamefont {King}},
  \bibinfo {author} {\bibfnamefont {Q.~A.}\ \bibnamefont {Turchette}}, \bibinfo
  {author} {\bibfnamefont {C.~A.}\ \bibnamefont {Sackett}}, \bibinfo {author}
  {\bibfnamefont {D.}~\bibnamefont {Kielpinski}}, \bibinfo {author}
  {\bibfnamefont {W.~M.}\ \bibnamefont {Itano}}, \bibinfo {author}
  {\bibfnamefont {C.}~\bibnamefont {Monroe}}, \ and\ \bibinfo {author}
  {\bibfnamefont {D.~J.}\ \bibnamefont {Wineland}},\ }\href {\doibase
  10.1038/35002001} {\bibfield  {journal} {\bibinfo  {journal} {Nature}\
  }\textbf {\bibinfo {volume} {403}},\ \bibinfo {pages} {269} (\bibinfo {year}
  {2000})}\BibitemShut {NoStop}%
\bibitem [{\citenamefont {Manzano}\ \emph {et~al.}(2019)\citenamefont
  {Manzano}, \citenamefont {Giorgi}, \citenamefont {Fazio},\ and\ \citenamefont
  {Zambrini}}]{Manzano2019}%
  \BibitemOpen
  \bibfield  {author} {\bibinfo {author} {\bibfnamefont {G.}~\bibnamefont
  {Manzano}}, \bibinfo {author} {\bibfnamefont {G.-L.}\ \bibnamefont {Giorgi}},
  \bibinfo {author} {\bibfnamefont {R.}~\bibnamefont {Fazio}}, \ and\ \bibinfo
  {author} {\bibfnamefont {R.}~\bibnamefont {Zambrini}},\ }\href {\doibase
  10.1088/1367-2630/ab5c58} {\bibfield  {journal} {\bibinfo  {journal} {New
  Journal of Physics}\ }\textbf {\bibinfo {volume} {21}},\ \bibinfo {pages}
  {123026} (\bibinfo {year} {2019})}\BibitemShut {NoStop}%
\bibitem [{\citenamefont {Breuer}\ and\ \citenamefont
  {Petruccione}(2002)}]{breuer2002}%
  \BibitemOpen
  \bibfield  {author} {\bibinfo {author} {\bibfnamefont {H.~P.}\ \bibnamefont
  {Breuer}}\ and\ \bibinfo {author} {\bibfnamefont {F.}~\bibnamefont
  {Petruccione}},\ }\href@noop {} {\emph {\bibinfo {title} {The Theory of Open
  Quantum Systems}}}\ (\bibinfo  {publisher} {Oxford University Press,
  Oxford},\ \bibinfo {year} {2002})\BibitemShut {NoStop}%
\bibitem [{\citenamefont {Rivas}\ and\ \citenamefont
  {Huelga}(2012)}]{rivas2012}%
  \BibitemOpen
  \bibfield  {author} {\bibinfo {author} {\bibfnamefont {A.}~\bibnamefont
  {Rivas}}\ and\ \bibinfo {author} {\bibfnamefont {S.~F.}\ \bibnamefont
  {Huelga}},\ }\href@noop {} {\emph {\bibinfo {title} {Open Quantum Systems: An
  Introduction}}}\ (\bibinfo  {publisher} {SpringerBriefs in Physics, Springer,
  Spain},\ \bibinfo {year} {2012})\BibitemShut {NoStop}%
\bibitem [{\citenamefont {Barra}(2015)}]{Barra2015}%
  \BibitemOpen
  \bibfield  {author} {\bibinfo {author} {\bibfnamefont {F.}~\bibnamefont
  {Barra}},\ }\href {\doibase 10.1038/srep14873} {\bibfield  {journal}
  {\bibinfo  {journal} {Scientific Reports}\ }\textbf {\bibinfo {volume} {5}},\
  \bibinfo {pages} {14873} (\bibinfo {year} {2015})}\BibitemShut {NoStop}%
\bibitem [{\citenamefont {Strasberg}\ \emph {et~al.}(2017)\citenamefont
  {Strasberg}, \citenamefont {Schaller}, \citenamefont {Brandes},\ and\
  \citenamefont {Esposito}}]{Strasberg2017}%
  \BibitemOpen
  \bibfield  {author} {\bibinfo {author} {\bibfnamefont {P.}~\bibnamefont
  {Strasberg}}, \bibinfo {author} {\bibfnamefont {G.}~\bibnamefont {Schaller}},
  \bibinfo {author} {\bibfnamefont {T.}~\bibnamefont {Brandes}}, \ and\
  \bibinfo {author} {\bibfnamefont {M.}~\bibnamefont {Esposito}},\ }\href
  {\doibase 10.1103/PhysRevX.7.021003} {\bibfield  {journal} {\bibinfo
  {journal} {Phys. Rev. X}\ }\textbf {\bibinfo {volume} {7}},\ \bibinfo {pages}
  {021003} (\bibinfo {year} {2017})}\BibitemShut {NoStop}%
\bibitem [{\citenamefont {Chiara}\ \emph {et~al.}(2018)\citenamefont {Chiara},
  \citenamefont {Landi}, \citenamefont {Hewgill}, \citenamefont {Reid},
  \citenamefont {Ferraro}, \citenamefont {Roncaglia},\ and\ \citenamefont
  {Antezza}}]{De_Chiara2018}%
  \BibitemOpen
  \bibfield  {author} {\bibinfo {author} {\bibfnamefont {G.~D.}\ \bibnamefont
  {Chiara}}, \bibinfo {author} {\bibfnamefont {G.}~\bibnamefont {Landi}},
  \bibinfo {author} {\bibfnamefont {A.}~\bibnamefont {Hewgill}}, \bibinfo
  {author} {\bibfnamefont {B.}~\bibnamefont {Reid}}, \bibinfo {author}
  {\bibfnamefont {A.}~\bibnamefont {Ferraro}}, \bibinfo {author} {\bibfnamefont
  {A.~J.}\ \bibnamefont {Roncaglia}}, \ and\ \bibinfo {author} {\bibfnamefont
  {M.}~\bibnamefont {Antezza}},\ }\href {\doibase 10.1088/1367-2630/aaecee}
  {\bibfield  {journal} {\bibinfo  {journal} {New Journal of Physics}\ }\textbf
  {\bibinfo {volume} {20}},\ \bibinfo {pages} {113024} (\bibinfo {year}
  {2018})}\BibitemShut {NoStop}%
\bibitem [{\citenamefont {Hewgill}\ \emph {et~al.}(2021)\citenamefont
  {Hewgill}, \citenamefont {De~Chiara},\ and\ \citenamefont
  {Imparato}}]{Hewgill2021}%
  \BibitemOpen
  \bibfield  {author} {\bibinfo {author} {\bibfnamefont {A.}~\bibnamefont
  {Hewgill}}, \bibinfo {author} {\bibfnamefont {G.}~\bibnamefont {De~Chiara}},
  \ and\ \bibinfo {author} {\bibfnamefont {A.}~\bibnamefont {Imparato}},\
  }\href {\doibase 10.1103/PhysRevResearch.3.013165} {\bibfield  {journal}
  {\bibinfo  {journal} {Phys. Rev. Research}\ }\textbf {\bibinfo {volume}
  {3}},\ \bibinfo {pages} {013165} (\bibinfo {year} {2021})}\BibitemShut
  {NoStop}%
\bibitem [{\citenamefont {Wichterich}\ \emph {et~al.}(2007)\citenamefont
  {Wichterich}, \citenamefont {Henrich}, \citenamefont {Breuer}, \citenamefont
  {Gemmer},\ and\ \citenamefont {Michel}}]{Wichterich2007}%
  \BibitemOpen
  \bibfield  {author} {\bibinfo {author} {\bibfnamefont {H.}~\bibnamefont
  {Wichterich}}, \bibinfo {author} {\bibfnamefont {M.~J.}\ \bibnamefont
  {Henrich}}, \bibinfo {author} {\bibfnamefont {H.-P.}\ \bibnamefont {Breuer}},
  \bibinfo {author} {\bibfnamefont {J.}~\bibnamefont {Gemmer}}, \ and\ \bibinfo
  {author} {\bibfnamefont {M.}~\bibnamefont {Michel}},\ }\href {\doibase
  10.1103/PhysRevE.76.031115} {\bibfield  {journal} {\bibinfo  {journal} {Phys.
  Rev. E}\ }\textbf {\bibinfo {volume} {76}},\ \bibinfo {pages} {031115}
  (\bibinfo {year} {2007})}\BibitemShut {NoStop}%
\bibitem [{\citenamefont {Horodecki}\ \emph {et~al.}(2009)\citenamefont
  {Horodecki}, \citenamefont {Horodecki}, \citenamefont {Horodecki},\ and\
  \citenamefont {Horodecki}}]{horodecki2009}%
  \BibitemOpen
  \bibfield  {author} {\bibinfo {author} {\bibfnamefont {R.}~\bibnamefont
  {Horodecki}}, \bibinfo {author} {\bibfnamefont {P.}~\bibnamefont
  {Horodecki}}, \bibinfo {author} {\bibfnamefont {M.}~\bibnamefont
  {Horodecki}}, \ and\ \bibinfo {author} {\bibfnamefont {K.}~\bibnamefont
  {Horodecki}},\ }\href {\doibase 10.1103/RevModPhys.81.865} {\bibfield
  {journal} {\bibinfo  {journal} {Rev. Mod. Phys.}\ }\textbf {\bibinfo {volume}
  {81}},\ \bibinfo {pages} {865} (\bibinfo {year} {2009})}\BibitemShut
  {NoStop}%
\bibitem [{\citenamefont {Peres}(1996)}]{peres1996}%
  \BibitemOpen
  \bibfield  {author} {\bibinfo {author} {\bibfnamefont {A.}~\bibnamefont
  {Peres}},\ }\href {\doibase 10.1103/PhysRevLett.77.1413} {\bibfield
  {journal} {\bibinfo  {journal} {Phys. Rev. Lett.}\ }\textbf {\bibinfo
  {volume} {77}},\ \bibinfo {pages} {1413} (\bibinfo {year}
  {1996})}\BibitemShut {NoStop}%
\bibitem [{\citenamefont {Horodecki}\ \emph {et~al.}(1996)\citenamefont
  {Horodecki}, \citenamefont {Horodecki},\ and\ \citenamefont
  {Horodecki}}]{horodecki1996}%
  \BibitemOpen
  \bibfield  {author} {\bibinfo {author} {\bibfnamefont {M.}~\bibnamefont
  {Horodecki}}, \bibinfo {author} {\bibfnamefont {P.}~\bibnamefont
  {Horodecki}}, \ and\ \bibinfo {author} {\bibfnamefont {R.}~\bibnamefont
  {Horodecki}},\ }\href {\doibase
  https://doi.org/10.1016/S0375-9601(96)00706-2} {\bibfield  {journal}
  {\bibinfo  {journal} {Phys. Lett. A}\ }\textbf {\bibinfo {volume} {223}},\
  \bibinfo {pages} {1 } (\bibinfo {year} {1996})}\BibitemShut {NoStop}%
\bibitem [{\citenamefont {Vidal}\ and\ \citenamefont
  {Werner}(2002)}]{vidal2002}%
  \BibitemOpen
  \bibfield  {author} {\bibinfo {author} {\bibfnamefont {G.}~\bibnamefont
  {Vidal}}\ and\ \bibinfo {author} {\bibfnamefont {R.~F.}\ \bibnamefont
  {Werner}},\ }\href {\doibase 10.1103/PhysRevA.65.032314} {\bibfield
  {journal} {\bibinfo  {journal} {Phys. Rev. A}\ }\textbf {\bibinfo {volume}
  {65}},\ \bibinfo {pages} {032314} (\bibinfo {year} {2002})}\BibitemShut
  {NoStop}%
\bibitem [{\citenamefont {Baumgratz}\ \emph {et~al.}(2014)\citenamefont
  {Baumgratz}, \citenamefont {Cramer},\ and\ \citenamefont
  {Plenio}}]{Baumgratz2014}%
  \BibitemOpen
  \bibfield  {author} {\bibinfo {author} {\bibfnamefont {T.}~\bibnamefont
  {Baumgratz}}, \bibinfo {author} {\bibfnamefont {M.}~\bibnamefont {Cramer}}, \
  and\ \bibinfo {author} {\bibfnamefont {M.~B.}\ \bibnamefont {Plenio}},\
  }\href {\doibase 10.1103/PhysRevLett.113.140401} {\bibfield  {journal}
  {\bibinfo  {journal} {Phys. Rev. Lett.}\ }\textbf {\bibinfo {volume} {113}},\
  \bibinfo {pages} {140401} (\bibinfo {year} {2014})}\BibitemShut {NoStop}%
\bibitem [{\citenamefont {Witten}()}]{Witten2018}%
  \BibitemOpen
  \bibfield  {author} {\bibinfo {author} {\bibfnamefont {E.}~\bibnamefont
  {Witten}},\ }\href {https://doi.org/10.1007/s40766-020-00004-5} {\bibinfo
  {journal} {arXiv:1805.11965}\ }\BibitemShut {NoStop}%
\end{thebibliography}%


%
\bibliographystyle{apsrev4-1}

\end{document}